\documentclass[twocolumn]{aastex63}
\usepackage[utf8]{inputenc}
\usepackage{gensymb}

\newcommand{\kms}{\mathrm{km~s^{-1}}}

\newcommand{\csm}{\mathrm{csm}}

\newcommand{\mdotvw}{\frac{\dot{M}}{v_w}}

\submitjournal{ApJ}

\shorttitle{Radio Observations of Young SNe Ia}
\shortauthors{Harris et al.}

\begin{document}

\title{Radio Observations of Six Young Type Ia Supernovae}

\correspondingauthor{C.\,E.~Harris}
\email{harr1561@msu.edu}

\author[0000-0002-1751-7474]{C.\,E.~Harris}
\affiliation{Department of Physics and Astronomy, Michigan State University, East Lansing, MI 48824, USA }

\author[0000-0002-4781-7291]{Sumit K. Sarbadhicary}
\affiliation{Department of Physics and Astronomy, Michigan State University, East Lansing, MI 48824, USA }
\affiliation{Department of Astronomy, The Ohio State University, 140 West 18th Avenue, Columbus, OH 43210, USA}
\affiliation{Center for Cosmology and Astroparticle Physics, The Ohio State University, 191 W. Woodruff Avenue, Columbus, OH 43210}

\author[0000-0002-8400-3705]{L.~Chomiuk}
\affiliation{Center for Data Intensive and Time Domain Astronomy, Department of Physics and Astronomy, Michigan State University, East Lansing, MI 48824, USA}

\author[0000-0001-6806-0673]{Anthony L. Piro}
\affiliation{The Observatories of the Carnegie Institution for Science, 813 Santa Barbara St., Pasadena, CA 91101, USA}

\author[0000-0003-4102-380X]{D. J. Sand}
\affiliation{Department of Astronomy/Steward Observatory, 933 North Cherry Avenue, Room N204, Tucson, AZ 85721-0065, USA}

\author[0000-0001-8818-0795]{S. Valenti}
\affiliation{Department of Physics, University of California, 1 Shields Avenue, Davis, CA 95616-5270, USA}

%
%
\begin{abstract}
    Type~Ia supernovae (SNe~Ia) are important cosmological tools, 
    probes of binary star evolution,
    and contributors to cosmic metal enrichment; 
    yet, a definitive understanding of the binary star systems that produce them
    remains elusive. 
    In this work we present early-time (first observation within 10 days post-explosion) 
    radio observations of six nearby (within 40 Mpc) SNe~Ia taken by
    the Jansky Very Large Array,
    which are used to constrain the presence of synchrotron emission
    from the interaction between ejecta and circumstellar material (CSM).
    The two motivations for these early-time observations are 
    (1) to constrain the presence of low-density winds and
    (2) to provide an additional avenue of investigation for those SNe~Ia
    observed to have early-time optical/UV excesses that may be due to
    CSM interaction.
    We detect no radio emission from any of our targets.
    Toward our first aim, these non-detections further increase the sample of SNe Ia that rule out winds from symbiotic binaries and strongly accreting white dwarfs.
    For the second aim, we present a radiation hydrodynamics simulation to explore
    radio emission from an SN~Ia interacting with a compact shell of CSM,
    and find that relativistic electrons cannot survive to produce radio emission
    despite the rapid expansion of the shocked shell after shock breakout. 
    The effects of model assumptions are discussed for both the wind and compact 
    shell conclusions.
\end{abstract}


\section{Introduction}\label{sec:intro}

    Type~Ia supernovae (SNe~Ia) are the most homogeneous type of stellar explosion, and
    are therefore crucial cosmological tools \citep{Branch98,Riess+16,DES19a,DES19b,Freedman+19}. 
    This homogeneity arises naturally as SNe~Ia are the complete explosion of carbon-oxygen
    white dwarfs (CO WDs) of similar mass \citep{WhelanIben73,Bloom+12,Churazov+15}.
    It is generally agreed that triggering an explosion requires the CO~WD to 
    interact with a binary companion star; however, the physical nature of this
    companion -- and, thus, the evolution to explosion -- is vigorously debated.
    Possible companions include  red giant (RG), sub-giant (SG), [post-]asymptotic 
    giant branch (AGB), main-sequence (MS), non-degenerate helium, He WD, or CO WD stars. 
    The remaining mysteries of SN~Ia progenitors are of general interest 
    because SNe~Ia impact a variety of topics in astrophysics, 
    such as studies of cosmology, 
    stellar evolution, next-generation gravitational wave signals, 
    and galaxy evolution.
    Overviews of the SN~Ia progenitor debate can be found in, e.g., \citet{Maoz+14} and 
    \citet{LivioMazzali18}.
    
    Constraining SN~Ia progenitors relies on indirect methods because, 
    unlike some massive star supernovae, 
    the expected progenitors are too dim to directly image 
    in the vast majority of cases.\footnote{SN~2011fe stands out as a notable exception, with constraints on
    giant and helium star companions made by \citet{Li+11c}.
    Thought to be non-terminal explosions, the ``SN~Iax'' class
    has also found some success with direct imaging techniques
    \citep[e.g.,][]{McCully+14}.}
    One such method is the study of the circumstellar medium (CSM) surrounding 
    the progenitor system. 
    For example SG and RG companions are all expected to drive stellar winds
    that will produce an extended CSM \citep{SeaquistTaylor90}. 
    MS/SG companions, as well as He~WD companions, that are accreting onto the CO~WD
    can produce CSM
    through non-conservative mass transfer \citep{HuangYu96}.
    High accretion rates can also drive an optically thick wind \citep{Hachisu+96}. 
    In any of these cases, we expect the wind to be hydrogen or helium rich,
    as the aforementioned accretion channels for SNe~Ia involve these elements.
    In addition to these wind-like CSM profiles, more complex environments have been
    theoretically posited, 
    such as a tidally ejected tail from the merger of two CO~WDs \citep{RaskinKasen13}, 
    the ejected common envelope from the merger of two CO~WDs or a CO~WD and 
    post-AGB star \citep{Hamuy+03,Soker13},
    or low-mass shells from a nova or recurrent novae \citep{MooreBildsten12}. 
    In these non-wind cases, the CSM may be confined rather than extended, 
    may lack hydrogen or helium,
    and its distance from the progenitor system at time of explosion is unknown.
    
    Constraining the circumstellar environment is primarily done by
    looking for signatures of the blast wave
    created by the collision of the SN~Ia ejecta with the CSM. 
    Shocks essentially redistribute the kinetic energy of the ejecta -- 
    donating some to ionizing the gas, heating the ions and electrons, 
    amplifying magnetic fields, accelerating ions and electrons to 
    relativistic speeds, and to radiation. 
    The exact mechanism by which this is accomplished is still
    an active area of research, with particular interest 
    in what share of the energy is given to each category and the
    downstream evolution of the components \citep[e.g.,][]{CaprioliSpitkovsky14}.
    Since shocks have the rare power to accelerate electrons 
    to relativistic speeds \textit{and} amplify magnetic fields, 
    one of the hallmark signatures of SN-CSM interaction is 
    radio emission from the synchrotron process.
    Although the interpretation of radio observations must then
    encounter the uncertainties in the shock physics related to 
    both the magnetic fields and relativistic electrons, 
    a key strength of this approach is that interaction with very low-density
    CSM is radio luminous although it would be invisible at most other 
    wavelengths.
    X-ray emission is similarly produced in low-density interactions
    and has the additional benefit of not relying on the 
    unknown physics of magnetic field amplification \citep{Margutti+12,Margutti+14}.
    However, radio observations have the key benefit of 
    being more sensitive and easier to mobilize -- as they are ground-based -- 
    such that it is possible to
    attain a larger sample of constraining radio observations than it is
    with X-ray instruments.
    Furthermore, interpreting X-ray emission can be complicated without a 
    spectrum to determine the relative contributions of thermal continuum,
    synchrotron continuum, inverse 
    Compton upscattering of the supernova photosphere, and 
    line emission \citep{ChevalierFransson06,Bochenek+18}.
    
    Given the benefits of radio observations to the SN~Ia progenitor
    puzzle, there is a host of literature on the topic.
    Primarily, studies have focused on the existence of a wind-like CSM
    due to the theoretical strengths of the symbiotic (RG companion) 
    progenitor channel in producing an explosion, which suggests that this
    channel should produce at least some SNe~Ia \citep{IbenTutukov84,Branch+95}.
    For example, a statistically significant set of radio observations
    near maximum light, assembled over decades, reveals that 
    the symbiotic channel is likely a minor player in the overall SN~Ia
    population \citep{Chomiuk+16}. 
    The CSM in this case has a density profile $\rho_\csm = qr^{-2}$, 
    with characteristic RG value $q \sim 5\times10^{11}~\mathrm{g~cm^{-1}}$. 
    In the cases of SN~2011fe and SN~2014J, deep radio limits probe even lower-density
    winds ($q \sim 10^{8}~\mathrm{g~cm^{-1}}$)
    and favor a double-degenerate merger scenario 
    \citep{Chomiuk2012,PerezTorres+14}.
    Overall, the emerging picture for typical SN~Ia progenitors 
    at radio frequencies is that
    the symbiotic channel is disfavored, and the question now is how to whittle 
    away at the remaining scenarios.
    
    Most radio observations of SNe Ia have been obtained on timescales 
    of weeks to a year after explosion \citep[e.g.][]{Chomiuk+16, Lundqvist+20}, 
    but here we argue two important reasons to search 
    for CSM interaction at radio frequencies within a week of explosion. 
    
    The first reason is that 
    interaction with low-density winds in other accretion scenarios, 
    such as novae or weak winds during the stable burning phase of the white dwarf, will 
    reach their \textit{peak} radio
    luminosity at these times, so early radio observations allow these winds to be constrained.
    
    The second reason is to reveal the nature of the early-time excess emission
    seen in some SNe~Ia \citep[e.g.,][]{Cao+15, Hosseinzadeh+17, Miller+18,Jiang+21,Deckers+22}. 
    There are many possible proposed origins for this emission, including
    interaction with CSM \citep{PiroMorozova16,Jiang+21},
    the impact of the ejecta with the companion \citep{Kasen10},
    He-shell detonation \citep[e.g.][]{Polin2019}, 
    radioactive Ni$^{56}$ mixing in the ejecta \citep[e.g.][]{PiroMorozova16, Magee2020, Sai+22},
    and the time evolution of the Dopper-shifted wavelength of
    prominent line absorption features (\citet{Ashall+22}; but see also \citet{Hosseinzadeh+22}).
    Estimates of the CSM mass and extent needed to explain early-excess
    SNe~Ia indicate dense, compact material. 
    \citet{Hosseinzadeh+17} model the early excess of SN~2017cbv
    as the interaction of an SN~Ia with a companion star but note that
    to obtain a similar signal with CSM interaction would require
    a CSM mass of $M_\csm \sim (0.01-0.1)~M_\odot$ extending to
    $R_\mathrm{out}\sim4\times10^{12}~\mathrm{cm}$,
    based on estimates given in \citet{Kasen10}, to significantly decelerate
    the ejecta.
    We note that the binary separation for the best-fit model to
    the data of SN~2012cg \citep{Marion+16} is similar to that
    found for SN~2017cbv, and would therefore require a similar
    CSM configuration. 
    For the case of SN~2020hvf, the rapid evolution of the early excess
    emission is well fit by a CSM interaction model where the 
    CSM has a mass $M_\csm=0.01~M_\odot$ extending to 
    $R_{\mathrm out}=10^{13}~\mathrm{cm}$ \citep{Jiang+21}. 
    For comparison, an RG wind would contain $\sim3\times10^{-9}~M_\odot$
    of material within this radius.
    The CSM in the early-excess case is much more dense than a typical wind, 
    yet radio emission is worth exploring because it is a ``smoking gun'' 
    signature of interaction.

    For these reasons, we have obtained deep 
    observations of
    six SNe~Ia within 10 days post-explosion 
    as part of several VLA programs, which we present
    in this manuscript and analyze in the context of the 
    low-density wind and compact shell CSM scenarios.
    In \S~\ref{sec:obs} we present the radio upper limits derived from our
    observations. 
    In \S~\ref{sec:models} we discuss our method for physically interpreting these limits;
    the vastly different density profiles between the low-density wind
    scenario (normal SNe~Ia) and the compact shell scenario (early-excess SNe~Ia)
    necessitates a separation in our modeling. 
    We first looking at the low-density winds, 
    for which we can use the \citet{Chevalier82a}
    self-similar hydrodynamic solution, 
    and show that this method cannot be applied to the compact shell 
    scenario, both because the shell is too dense and because it is truncated 
    rather than extended (\S~\ref{ssec:wind_hydro}). 
    We then describe how we use numerical radiation hydrodynamics to simulate
    the evolution of the shock in the compact CSM case in \S~\ref{ssec:shell_hydro}.
    In \S~\ref{ssec:electrons} we take a careful look at the relativistic
    electron populations in the shock region that are needed to create synchrotron
    emission, since we are concerned about whether these survive at all in
    the compact shell interaction, and discuss unknown physical parameters that
    affect radio analysis.
    \S~\ref{ssec:model_lc} describes the synchrotron light curve calculation 
    for the self-similar solution.
    In \S~\ref{sec:results} we then present our analysis of the radio limits
    in the context of our models, showing how our sample significantly increases
    the number of SNe~Ia that constrain low-density winds and taking a detailed
    look at the possibility of radio emission from early-excess SNe~Ia.
    A summary of the work can be found in \S~\ref{sec:conc}.

\section{Early-Time Observations of Type Ia Supernovae}\label{sec:obs}

\begin{deluxetable*}{lclrll}
 \tablecaption{Summary of the young SNe Ia sample}
 \label{tab:obs}
  \tablehead{Name & Location$^{a}$ & Host galaxy$^{a}$ & D (Mpc)$^{b}$ & Discovery Date$^{a}$ & Last Non-detection$^{a,e}$}
  \startdata
SN 2019np & 10:29:21.980 +29:30:38.30 & NGC 3254 & 19.4 & 2019 Jan-09.66 & 2019 Jan-08.53 \\
SN 2019ein$^{c}$ & 13:53:29.110 +40:16:31.33 & NGC 5353 & 33 & 2019 May-01.47 & 2019 April-29.27 \\
SN 2020rcq & 11:57:14.680 +49:17:31.99 & UGC 6930 & 11.1 & 2020 Aug-09.17 & 2020 Aug-07.17 \\
SN 2020uxz & 01:24:06.890 +12:55:17.26 & NGC 514 & 35.5 & 2020 Oct-05.57 & 2020 Oct-05.40 \\
SN 2021qvv$^{d}$ & 12:28:02.920 +09:48:10.26 & NGC 4442 & 14.7 & 2021 June-23.26 & 2021 June 21.37 \\
SN 2021smj & 12:26:46.560 +08:52:57.61 & NGC 4411b & 22.4 & 2021 July-08.27 & 2021 July-5.28 \\
\enddata
\tablecomments{\footnotesize $^{a}$Obtained from Transient Name Server (TNS)\\
$^{b}$Distances measured from corresponding redshifts of the SN listed in TNS using the Cosmology Calculator \citep{cosmologycalculator},assuming $H_0 = 70$ $\mathrm{km/s/Mpc}$, $\Omega_{M}=0.3$ and $\Omega_{\Lambda}=0.7$.\\
$^{c}$Measurements obtained from \cite{Pellegrino2020}. Exhibits high-velocity features.\\
$^{d}$91bg-like, according to TNS\\
$^{e}$Since most SNe on TNS have photometry reported by multiple groups, we use the epoch of last non-detection reported by the discovery group on TNS, and if not available, the last non-detection by any of the other groups.}
\end{deluxetable*}

\begin{deluxetable*}{llrlcrrr}
  \tablecaption{Summary of radio observations of SNe Ia}
 \label{tab:obs_lims}
  \tablehead{Name & Project & Config$^{a}$ & Observation  & Synthesized Beam & Age$^{b}$ & F$^{upper}_{6\mathrm{GHz}}$ (3$\sigma$)$^{c}$ & L$^{upper}_{6\mathrm{GHz}}$ (3$\sigma$)$^{d}$\\
  & Code & & Date & (arcsec$\times$arcsec) & (days) & ($\mu$Jy) & ($\times 10^{24}$ ergs/s/Hz)}
\startdata
SN 2019np & 18B-162 & C & 2019 Jan-11.36  & 2.78 $\times$ 2.64 & 2.8 & 19.7 & 8.9 \\
 & & & 2019 Jan-18.35  & 2.73 $\times$ 2.49 & 9.8 & 19.3 & 8.7\\
 & & & 2019 Jan-25.45  & 2.88 $\times$ 2.48 & 16.9 & 15.6 & 7.0\\
 \hline
SN 2019ein & 19A-010 & B & 2019 May-03.35  & 1.28$\times$1.27 & 3.9 & 17.9 & 23.4 \\
            & & & 2019 May-11.04  & 2.20$\times$1.23 & 11.6 & 25.3 & 33.1 \\
            & & & 2019 May-17.06  & 1.77$\times$1.99 & 17.6 & 23.5 & 31.0 \\
\hline
SN 2020rcq & 20B-355 & B & 2020 Aug-16.81  & 1.02 $\times$ 0.73 & 9.6 & 17.2 & 2.5 \\
& & & 2020 Aug-23.73  & 1.09$\times$0.74 & 16.6 & 17.0 & 2.5 \\
& & & 2020 Aug-25.8  & 0.94$\times$0.74 & 18.6 & 18.9 & 2.8 \\
\hline
SN 2020uxz & 20B-355 & B & 2020 Oct-06.28  & 0.87$\times$0.74 & 0.9 & 18.9 & 28.6 \\
& & & 2020 Oct-16.41  & 1.19$\times$0.87 & 11 & 18.1 & 27.4 \\
& & & 2020 Nov-05.28  & 0.90$\times$0.36 & 30.9 & 17.9 & 27.0 \\
\hline
SN 2021qvv & 21B-295 & C & 2021 June-24.99  & 3.18$\times$2.7 & 3.6 & 21.6 & 5.6\\
& & & 2021 July-09.82  & 5.4$\times$2.89 & 18.5 & 28.0 & 7.3 \\
& & & 2021 July-16.99  & 3.07$\times$2.49 & 25.6 & 20.6 & 5.3 \\
\hline
SN 2021smj & 21B-295 & C & 2021 July-10.88  & 3.57$\times$3.08 & 5.6 & 18.1 & 10.9 \\
& & & 2021 July-15.99  & 3.17$\times$2.56 & 10.7 & 17.3 & 10.4 \\
\enddata
\tablecomments{\footnotesize $^{a}$ VLA Configuration of observation.\\
$^{b}$Since last non-detection in Table \ref{tab:obs}.\\
$^{c}$3$\sigma$ upper limit to the 6 GHz flux density, described in Section \ref{sec:obs}.\\
$^{d}$3$\sigma$ upper limit to the 6 GHz spectral luminosity.}
\end{deluxetable*}

In this section, we describe our methodology for obtaining 6 GHz radio
luminosity limits for our target SNe~Ia.
The targets and their limits are given in Tables~\ref{tab:obs} \& \ref{tab:obs_lims}.

We obtained VLA observations of young Type Ia SNe\footnote{Due to the 
declination limit of VLA, we were only able to observe targets above a 
declination of $-40^{\degree}$ with our program.} for the last few years 
within 40 Mpc (more distant SNe will not have deep enough constraints in radio). 
We used target-of-opportunity VLA programs to detect potential radio emission 
promptly after discovery as expected from main-sequence companions or close-in 
CSM shells. 
Once a SN was posted and spectroscopically classified as a Type Ia on 
the Transient Name Server (TNS), we triggered a VLA observation, along with 
typically 1--2 follow-up observations scheduled a week apart. 
Each observing block was 1 hr long in C-band (4--8 GHz), corresponding to an 
image RMS of roughly 4 $\mu$Jy/beam or a 3$\sigma$ luminosity limit of 
roughly $2.6 \times 10^{25}$ ergs/s/Hz at 40 Mpc, enough to detect dense 
shells and winds predicted in SN Ia progenitor models \citep{Chomiuk2012}. 
Although the synchrotron emission from SNe will be brighter in L-band (1-2 GHz), 
we chose to observe in C-band due to its larger instantaneous bandwidth 
($\sim$3.4 GHz, excluding RFI) and lower system temperature, giving 
increased sensitivity to faint sources for the same integration time. 
Moreover, C-band observations provide higher angular resolution images 
than L-band at any array configuration, which helps to distinguish emission 
from the SN location from other nearby confusing sources (e.g. bright galactic nuclei). 

Our sample consists of six SNe newly observed with our VLA program between 2018-2021.
With the exception of SN2019ein, whose radio observations have been published elsewhere \citep{Pellegrino2020}, we calibrated, imaged and recorded radio flux densities 
and upper limits of our targets in this paper.

Calibration was done with CASA pipeline versions 5.6--6.1. 
Aside from minor differences in implementation between versions, 
the pipeline applies online flags reported by the NRAO operator
(e.g. antenna position changes) and performs iterative flagging and calibration
of the raw visibility data. Calibration consists of delay, bandpass and absolute
flux density scale calibration with one of VLA’s primary calibrator, and antenna-based 
complex gain calibration using a secondary calibrator close to the target location 
and observed in between target observations. 
Flagging is done with the automated rflag algorithm in between calibration steps 
to remove corrupted measurements. Each pipeline calibrated dataset is manually 
inspected and further flagged if RFI remains, and then passed on to imaging.

Imaging was performed using the \texttt{tclean} task. The imaging region extended 
to a primary beam level of 5$\%$ to ensure all outlying sources were deconvolved 
in order to obtain maximum sensitivity at the image center where the SN target is located. 
The cell size was chosen to sample the full width at half maximum of the effective point-spread-function (PSF) with 4 pixels. 
For all images, we used \texttt{gridder=standard}, 
which resamples the visibility data onto a regular 
grid using a prolate spheroidal function. Gridding options that include widefield, direction-dependent corrections exist in CASA, but are computationally expensive, 
and likely unnecessary since our object of interest is at the phase center. 
For deconvolution, we use multi-term multi-frequency synthesis with \texttt{nterms=2} to account 
for the frequency-dependent structure of the sky. Briggs weighting with \texttt{robust=0} was 
applied to the data in order to balance point source sensitivity with sidelobe 
contamination from bright sources. 
With these settings, we cleaned our images for a maximum of 10$^4$ iterations, 
or once the image RMS is less than three times the peak residual 
in the image (\texttt{nsigma=3}). 
In cases where the RMS of the final image is limited by dynamic range due to 
bright sources, we carried out phase-only self-calibration to further improve the image.

No SN was detected in any of our radio images. We provide the results of our observations in Table \ref{tab:obs_lims}. The 6 GHz 3$\sigma$ upper limit on the flux density for each SN is defined as the 6 GHz flux density at the pixel location of the SN, plus the 3$\sigma$ RMS noise calculated in a circular region around the SN. In the case where the flux density at the pixel location is negative, we only use the 3$\sigma$ RMS noise. The radii of the regions ranged from 6-14$^{\prime\prime}$, as they were adjusted to be large enough to contain sufficient pixels for a robust RMS calculation, but small enough to represent the local RMS and exclude any visible adjacent sources or artifacts. The 6 GHz 3$\sigma$ luminosity is then given by $\left(1.2 \times 10^{21} \mathrm{ergs\ s^{-1}\ Hz^{-1}}\right) S_{\mu jy} D_{Mpc}^2$ as per Eq. 3 in \cite{Chomiuk09}, where $S_{\mu jy}$ is the 3$\sigma$ flux upper limit in $\mu$Jy, and $D_{Mpc}$ is the distance to the SN in Mpc, given in Table \ref{tab:obs}.

One object in our sample, SN~2019np, is observed
to have a modest early excess in the bolometric light curve, but
this excess is well explained by $^{56}$Ni mixing in the ejecta
\citep{Sai+22}.

\section{Ejecta-CSM Interaction Models}\label{sec:models}

\begin{figure*}
        \centering
        \includegraphics[width=.9\textwidth]{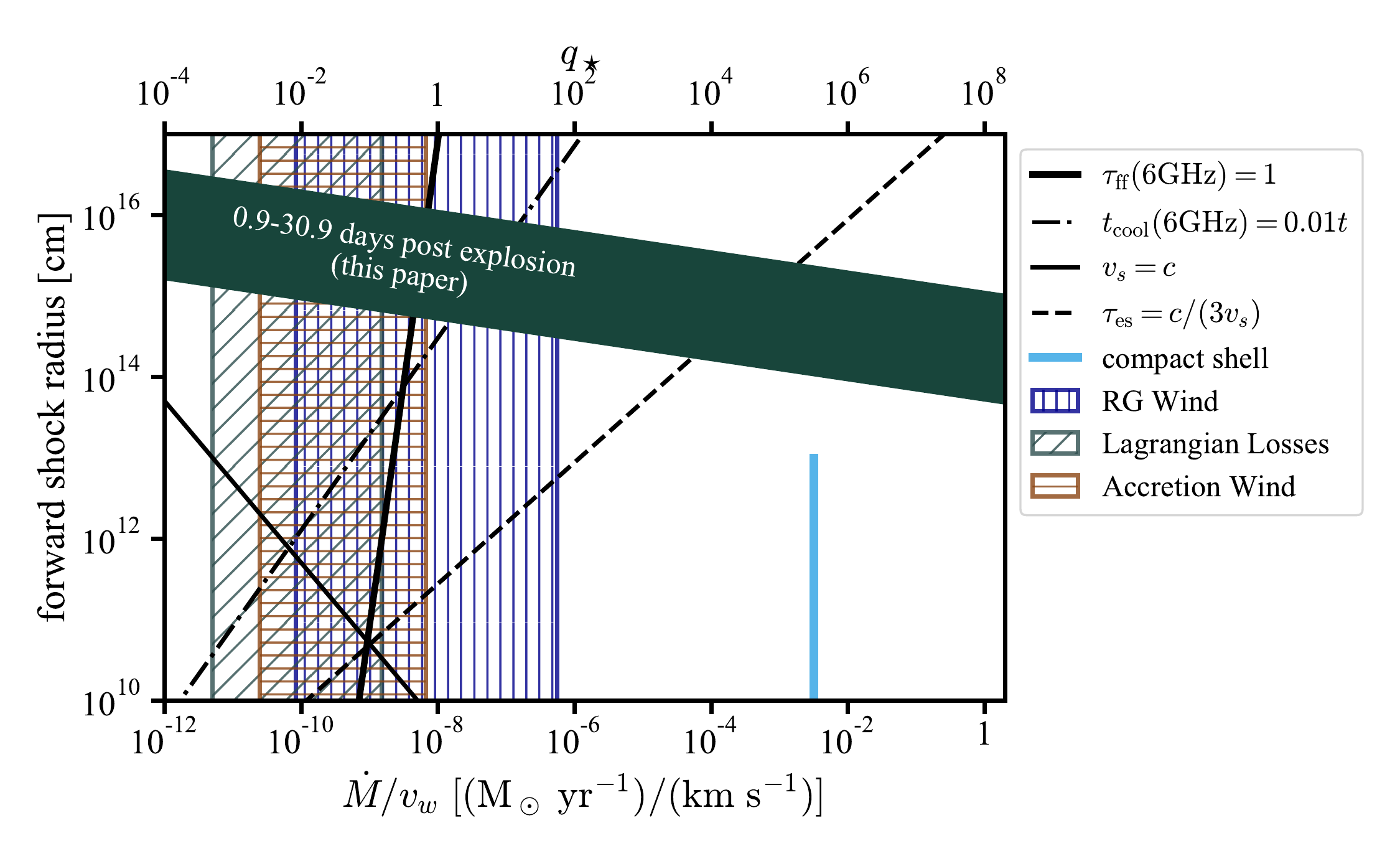}
        \caption{Quantities of interest for our analysis, assuming 
        the forward shock radius evolves according to the self-similar
        solution for interaction with a wind.
        The shock radii probed by our observations are shown by the 
        dark green band, hatched regions show wind CSM
        configurations we investigate, and 
        the thick blue line describes our compact shell simulation.
        The assumptions of the hydrodynamic model are valid for $R_f$
        above the $v_s=c$ (thin solid) and $\tau_\mathrm{es}=c/(3v_s)$ (dashed) lines.
        The $\tau_\mathrm{ff}$ (thick solid) line shows where an infinite wind 
        becomes optically thin to external free-free absorption -- right of this
        line, 6 GHz emission is absorbed.
        The $t_\mathrm{cool}=0.01t$ (dot-dashed) line shows the relativistic
        electron cooling boundary -- right of this line, the emitting volume is reduced.
        }
        \label{fig:selfsim_applicability}
    \end{figure*}

    In this section, we describe the hydrodynamic and radio synchrotron 
    light curve models used in this paper to analyze the early-time
    radio observations.
    \S~\ref{ssec:wind_hydro} describes the analytic hydrodynamic
    solution we employ for constraining the presence of low-density
    winds, and \S~\ref{ssec:shell_hydro} describes our 
    radiation hydrodynamics simulation for approaching the 
    compact shell scenario.
    We review the model for the calculation of the 
    electron acceleration and synchrotron light curves that result from the 
    interaction in \S~\ref{ssec:electrons} \& \ref{ssec:model_lc}. 
    Figure~\ref{fig:selfsim_applicability} provides
    a visual synopsis of this section, showing the shock radii
    probed by our observations, the CSM configurations of interest,
    and the various limiting factors for our analysis. 
    In \S~\ref{sec:results}, we will use the methodology presented here to
    physically interpret our radio limits.

    \subsection{Wind Interaction Hydrodynamics}\label{ssec:wind_hydro}
        For the interaction with steady winds, we employ the 
        self-similar model of \citet{Chevalier82a}. 
        For interaction with a wind-like CSM, this model has performed
        well in fitting the radio data of core-collapse SNe \citep[e.g.,][]{Weiler+86,Soderberg+05}.
    
        \subsubsection{Ejecta Density Profile for Steady Wind Interaction}\label{ssec:wind_ejecta_dens}
            For wind interaction, we will approximate the SN~Ia ejecta
            as a broken power-law 
            \citep{ChevalierFransson94}, which provides a reasonable approximation to 
            the structure found in hydrodynamical models of SNe. 
            As in \citet{Kasen10}, we use $\rho \propto r^{-1}$
            to describe the inner ejecta mass-density profile. 
            We note that the detailed structure of the inner ejecta is 
            only relevant in this work in setting the fraction of the total
            mass that is in the outer ejecta. 
            We assume the
            ejecta are in homologous expansion such that $v = r/t$,
            with $v$ being the ejecta speed and $t$ time.
            The transition point  between the inner and outer ejecta occurs
            at speed 
            \begin{equation}
                v_t = (1.014 \times 10^{4}~ \kms)~ (E_{51}/M_\mathrm{c})^{1/2},           
                \label{eqn:v_t}            
            \end{equation}
            where $M_\mathrm{c} = M_\mathrm{ej}/(1.44~M_\odot)$ is ejecta mass in units of the 
            Chandrasekhar mass and $E_{51} = E/(10^{51}~ \mathrm{erg})$ is the explosion energy.
            We use $M_c=1$ and $E_{51}=1$. 
            The outer ejecta density profile is given by 
            \begin{eqnarray}
                \rho_\mathrm{ej}(r) = \frac{0.124 M_\mathrm{ej}}{(v_t t)^3} 
                \left( \frac{r}{v_t t} \right)^{-10}, \label{eqn:rhoej_orig}
            \end{eqnarray}
            which can be written as 
            \begin{eqnarray}
                \rho_\mathrm{ej} &=& \rho_\mathrm{0,ej} t^7 r^{-10} , \label{eqn:rhoej}\\
                \rho_\mathrm{0,ej} &\equiv& (3.6\times10^{9}~\mathrm{g~cm^{-3}})^{10} M_c^{-5/2} E_{51}^{7/2} \label{eqn:rhoej_norm} ~.
            \end{eqnarray}
            
        The above equations describe the density and velocity profile for 
        the SN~Ia ejecta we assume when using the self-similar model 
        for the wind interaction case.

        \subsubsection{Density Profile of Steady Winds}\label{sssec:wind_dens}
        
        A steady wind can be described by the density profile
        $\rho = qr^{-2}$. 
        For consistency with \citet{ChevalierFransson06} and 
        \citet{Chomiuk+16}, we will adopt the normalization 
        $q_* = q/(5\times10^{11}~\mathrm{g~cm^{-1}})$,
        such that $q_*=1$
        corresponds to $\dot{M}/v_w = 10^{-8}~\mathrm{(M_\odot~yr^{-1})/(km~s^{-1})}$ --
        though we note that this variable is called ``$A_*$'' in those works. 
        Thus,
        \begin{eqnarray}
                \rho_\csm = (5\times10^{-19}~\mathrm{g~cm^{-3}}) q_\star \left( \frac{r}{10^{15}~\mathrm{cm}}\right)^{-2}
        \end{eqnarray}

        Although the self-similar model constrains $q$ without 
        reference to the wind origin, throughout this work we will 
        reference several physically motivated winds.
        First, there is the canonical symbiotic system in the single-degenerate
        scenario, with a CO~WD accreting a RG wind. 
        For this, we assume wind velocities span $v_w \sim (9-60)~\mathrm{km~s^{-1}}$
        and mass-loss rates span 
        $\dot{M} \sim (5\times10^{-9} - 5\times10^{-6})~\mathrm{M_\odot~yr^{-1}}$
        \citep{SeaquistTaylor90}. 
        Second, we consider the properties of an optically thick wind
        from high accretion rates \citep{Hachisu+96}, 
        for which we use a velocity span of 
        $v_w \sim (300-4,000)~\mathrm{km~s^{-1}}$
        and consider mass-loss rates 
        $\dot{M} \sim (10^{-7} - 2\times10^{-6})~\mathrm{M_\odot~yr^{-1}}$.
        These winds could be in play in either hydrogen or helium accretion
        scenarios \citep{Hachisu+99}, but this mass-loss scenario may not
        be in effect at the time of explosion as it requires a high accretion rate.
        Finally, we show the hypothetical case of non-conservative mass
        transfer via losses from the outer Lagrangian point in a 
        Roche Lobe overflow scenario \citep{HuangYu96}.
        Here, we assume the accretion rate is appropriate for stable burning
        \cite[e.g.,][]{ShenBildsten07}
        and the mass-loss rate is a modest 1\% of the accretion rate,
        such that 
        $v_w \sim (20-600)~\mathrm{km~s^{-1}}$
        and 
        $\dot{M} \sim (10^{-9} - 3\times10^{-8})~\mathrm{M_\odot~yr^{-1}}$.
        These winds are shown in Figure~\ref{fig:selfsim_applicability}
        as the ``RG wind,'' ``Accretion Wind,'' and ``Lagrangian losses'' 
        regions, respectively.

    \subsubsection{Steady Wind Shock Hydrodynamics}\label{sssec:wind_hydro}
        We will leverage the asymptotic 
        self-similar solution of \citet{Chevalier82a}
        for the hydrodynamics of the shock evolution. 
        Per this model, the contact discontinuity radius (the boundary
        between the ejecta and CSM) evolves as
        \begin{eqnarray}
            R_c = \left(\frac{A\rho_\mathrm{0,ej}}{q}\right)^{1/(n-s)} t^{(n-3)/(n-s)} ~,
        \end{eqnarray}
        where $\rho_\mathrm{ej} = g^n t^{n-3} r^{-n}$, $\rho_\mathrm{csm}=qr^{-s}$, 
        $t$ is time, and $A$ is a numerical constant that depends on $s$ and $n$. 
        From our ejecta and CSM density profiles, we identify
        $g^n = \rho_\mathrm{0,ej}$, 
        $n=10$, and $s=2$, such that $A=0.067$,  $R_f/R_c = 1.239$, 
        $R_r/R_c = 0.984$, and thus, 
        \begin{eqnarray}
            R_{f,15} = 4.22~M_c^{-5/16} E_{51}^{7/16} q_\star^{-1/8} t_{10}^{7/8}~,
            \label{eqn:shock_R}
        \end{eqnarray}
        where $R_{f,15} = R_f/(10^{15}~\mathrm{cm})$.    
    
        Since our work involves very early-time observations and 
        we consider interaction with dense CSM, 
        we will here revisit two underlying assumptions of the model 
        that can affect the shock hydrodynamics.

        \paragraph{Assumption 1: Non-relativistic Shock}
        As can be seen in Equation \ref{eqn:shock_R}, the shock is
        decelerating over time, and the speed of the shock becomes
        infinite at $t=0$. 
        We must ensure that our early-time observations are not 
        taken at a time when this model would imply a 
        relativistic shock, since the model is non-relativistic.
        With the equation for $R_c$, time can be written
        \begin{eqnarray}
            t_{10} &=& 0.193~M_c^{5/14} E_{51}^{-1/2} R_{f,15}^{8/7} q_*^{1/7}  ,
        \end{eqnarray}
        where $t_{10} = t/(10~\mathrm{days})$.
        The forward shock speed is $v_s = [(n-3)/(n-s)] R_f t^{-1}$, so
        \begin{eqnarray}
            v_{s,9} &=& 5.25 (M_c^{5/2} E_{51}^{-7/2} q_* R_{f,15})^{-1/7}  
            \label{eqn:v_s_Rfunc}\\
            &=& 4.28 (M_c^{5/2} E_{51}^{-7/2} q_\star t_{10})^{-1/8}
            \label{eqn:v_s_tfunc}~,
        \end{eqnarray}
        where $v_{s,9} = v_s/(10^9~\mathrm{cm~s^{-1}})$.
        Then, the constraint that $v_s$ is less than the speed of light, $c$, can be 
        written as the constraint that the forward shock radius is 
        \begin{eqnarray}
            R_{f,15} \geq (5.06\times 10^{-6})~M_c^{-5/2} E_{51}^{7/2} q_\star^{-1} ~,
        \end{eqnarray}
        where $R_{f,15} = R_f/(10^{15}~\mathrm{cm})$.
        This constraint is shown in Figure~\ref{fig:selfsim_applicability}
        and is not a concern for our analysis. 
        
        As we are discussing shock speed, we note that the ram pressure of the shock is 
        \begin{eqnarray}
        \frac{\rho_\csm v_s^2}{\mathrm{erg~cm^{-3}}} &=& 13.79 M_c^{-5/7} E_{51} q_*^{5/7} R_{f,15}^{-16/7} \label{eqn:rampress_r}\\
        &=& 0.51 q_* t_{10}^{-2} \label{eqn:rampress_t} ~,
        \end{eqnarray}
        which will be used to determine post-shock energy densities.

        \paragraph{Assumption 2: Equation of State}
        The model we are using assumes 
        a gamma-law equation of state with $\gamma_\mathrm{ad}=5/3$, but 
        the high densities of some of the CSM configurations of interest may 
        be radiaton pressure dominated ($\gamma_\mathrm{ad}=4/3$), which would
        violate this assumption. 
        We can estimate the boundary 
        as in \citet{ChevalierFransson94}, 
        arguing that photons are not trapped by the gas so long as the
        electron scattering optical depth obeys
        \begin{equation}
            \tau_\mathrm{es} < c/(3v_s)~.
        \end{equation}
        Using
        \begin{eqnarray}
            \tau_\mathrm{es} = \kappa_\mathrm{es} (4\rho_\csm) (1 - R_r/R_f) R_f ,
        \end{eqnarray}
        where the electron scattering opacity is 
        $\kappa_\mathrm{es} = 0.4~\mathrm{cm^2~g^{-1}}$, we arrive at the 
        condition that photons can escape the shock region when
        \begin{eqnarray}
            R_f &>& \left(\mdotvw\right)^{\frac{n-4}{n-2}} 
            (4\pi A \rho_\mathrm{0,ej})^{\frac{1}{n-2}} \\
            &\times& \left(\frac{R_f}{R_c}\right)
            \left(1-\frac{R_r}{R_f}\right)^{\frac{n-3}{n-2}} 
            \left(\frac{(n-3)(3\kappa_\mathrm{es})}{(n-2)(\pi c)}\right)^{\frac{n-3}{n-2}} .
        \end{eqnarray}
        For the ejecta and CSM used in this work, this condition is
        \begin{eqnarray}
            R_f > (2.81\times10^{11}~\mathrm{cm})~q_\star^{3/4}~. \label{eqn:rad_dom_cond}
        \end{eqnarray}
        This constraint is shown in Figure~\ref{fig:selfsim_applicability}, 
        where we see that radiation hydrodynamics calculations are
        required for the highest densities.
        As mentioned in \S~\ref{sec:intro},
        compact shells must have $q_\star\gtrsim10^5$ to explain early-excess 
        SNe~Ia and are clearly radiation pressure dominated, 
        necessitating numerical radiation hydrodynamic simulations to 
        capture their shock evolution.

    \subsection{Compact Shell Interaction Radiation Hydrodynamics}\label{ssec:shell_hydro}
        As shown in the last section,
        the CSM shells expected to produce optical excess from interaction 
        are in a regime where shocks are radiation 
        dominated.
        Therefore, to model the interaction of SNe~Ia with compact shells of 
        CSM, we  
        use the SuperNova Explosion Code (SNEC), a one-dimensional Lagrangian 
        radiation hydrodynamics code \citep{Morozova+15} to extend the set of
        compact CSM interaction models presented in \citet{PiroMorozova16}.
        We refer the reader to these works for further detail, but we
        summarize key points of the modeling below.
        
        There are several reasons to expect that we would not detect radio 
        emission during the interaction itself.
        First, because the CSM is truncated at a very small radius 
        ($\sim10^{13}~\mathrm{cm}$) so a radio observation would have to be
        obtained within about an hour post-explosion to observe at this phase.
        Second, the CSM is so dense that radio emission will be easily 
        absorbed by the pre-shock CSM 
        (see \S~\ref{ssec:model_lc} for a calculation).
        However, in the case of adiabatic shocks with truncated CSM, 
        \citet{HNK16} showed that there is a long tail to the radio
        light curve after the shock has crossed the CSM.
        Numerical simulations are required to capture the evolution of the
        shocked CSM as it rarefies to assess whether such a tail 
        could be observed in the compact shell interaction case.
        
        \subsubsection{Ejecta Density Profile for Compact Shell Interaction}\label{ssec:shell_ejecta_dens}    
        
        The WD explodes as part of the SNEC simulation, so the 
        ejecta density profile in this case is generated within the
        simulation. 
        The WD density profile is that of a $1.25~M_\odot$ WD evolved 
        with the Modules for Experiments in Stellar Astrophysics (MESA) code
        \citep{Paxton+11}.
        Its composition is 
        49\% $^{12}$C, 49\% $^{16}$O, and 2\% $^{20}$Ne by mass. 
        Since SNEC lacks a nuclear reaction network, the composition
        does not change post-explosion, except that $^{56}$Ni is added
        to the domain.
        As in \citet{PiroMorozova16}, we input 
        $0.5~M_\odot$ of $^{56}$Ni 
        (the mass fractions of other elements are adjusted to account for this).
        For the extent of the $^{56}$Ni distribution, we use a 
        boxcar width of $0.125~M_\odot$, which matches the $V$-band 
        rise of SN~2011fe best of the Ni-56 distributions explored in
        \citet{PiroMorozova16}. 
        More distributed nickel causes a shallower light curve rise,
        as seen in SN~2019np \citep{Sai+22}.

        \subsubsection{Compact Shell Density Profiles}
    
        Motivated by several WD merger models \citep{Pakmor+12,Schwab+12,Shen+12},
        the compact shell interaction models produced by 
        \citet{PiroMorozova16} have 
         $M_\csm = 0.1$ and $R_\mathrm{out}=(10^9 - 10^{12})~\mathrm{cm}$,
        and a density profile 
        \begin{equation}
            \rho_\mathrm{shell} = q_\mathrm{shell} r^{-3} ~,
        \end{equation}
        where $q_\mathrm{shell}$ can be found via
        $M_\csm = 4\pi q_\mathrm{shell} \ln (R_\mathrm{out}/R_\mathrm{WD})$,
        with $R_\mathrm{WD}$ the white dwarf outer radius. 
        As it is supposed to have originated from the merger of two WDs, we 
        use the same composition for the CSM as for the WD zones
        (49\% $^{12}$C, 49\% $^{16}$O, 2\% $^{20}$Ne).
        In this work, we present a model with 
        $M_\csm = 0.01~M_\odot$ and $R_\mathrm{out} = 10^{13}~\mathrm{cm}$,        
        which describes the CSM of the early-excess SN~2020hvf \citep{Jiang+21}.
        Our simulation differs from that of \citet{Jiang+21}, however,
        because we assume normal SN~Ia ejecta properties
        and model the first day post-explosion with increased time sampling
        (since our focus is the evolution of the shock). 
        This model is illustrated in
        Figure~\ref{fig:selfsim_applicability},
        using $\dot{M}/v_w = M_\csm/R_\mathrm{out}$.

        \subsubsection{Radiation Hydrodynamics with SNEC}\label{sssec:shell_shock_hydro}
        
        In the compact shell CSM case, the high densities imply that the 
        shock is radiation pressure dominated (Equation~\ref{eqn:rad_dom_cond}). 
        At the same time, these compact shells are of finite extent, 
        so there will be a shock breakout (the shocked gas becomes optically thin)
        near the time that the shock front reaches the edge of the CSM, 
        after which point the system becomes gas pressure supported.
        Due to this complex behavior and the necessity of following the evolution
        of the shocked gas after breakout, we employ the SNEC radiation 
        hydrodynamics code to explore this scenario.
        As in \citet{PiroMorozova16}, our simulations use piston-driven explosions to
        unbind the WD.
        The domain is 400 zones covering the entire WD and CSM, with increased
        mass resolution at the inner and outermost zones. 
        The output includes the evolution of hydrodynamic variables in each zone, 
        which we can then use in calculations of interest. 
        Furthermore, SNEC output provides the evolution of the shock speed and 
        radius which we utilize to identify the immediate postshock zones.

    \subsection{Electron Distribution}\label{ssec:electrons}
    
    In any interaction scenario, radio emission through the synchrotron process
    depends on relativistic electrons.
    Synchrotron emission arises from electrons accelerated to 
    relativistic speeds through repeated interactions with 
    the amplified and turbulent magnetic field in the shocked gas
    \citep[see, e.g.,][for a review]{Marcowith+20}. 
    Then, for Lorentz factors between $\gamma_{\min}$ and $\gamma_{\max}$, 
    the electron density is described by a power-law distribution
    \begin{eqnarray}
        N_{e,\mathrm{nt}}(\gamma) d\gamma = C_\gamma \gamma^{-p} d\gamma,
    \end{eqnarray}
    where $n_{e,\mathrm{nt}} = \int N_{e,\mathrm{nt}} d\gamma$ is the 
    number density of electrons in the power-law distribution.
    Therefore, to describe the relativistic electrons that will create 
    synchrotron emission, we must determine $C_\gamma$, 
    $\gamma_{\min}$, $\gamma_{\max}$, and $p$. 
    In this work, we follow the method of \citet[][]{ChevalierFransson06}
    for accomplishing this, as we detail below.

    The value of $p$ is expected to be $2<p<3$. 
    \citet{Chevalier98} takes $p=3$ as a reference value based on observations of 
    SNe~Ibc. 
    In other contexts, the index may differ slightly; for example,
    observations of supernova remnants (SNRs) have found $p=2.2-2.5$ \citep{Green+19}. 
    \citet{DiesingCaprioli21} suggest that the steeper power-law index 
    in radio SNe may be explained by their faster shocks
    and higher magnetic field energy densities.
    In this work, we will use $p=3$ as our fiducial value because
    we expect the shock speeds to be more similar to SNe~Ibc than
    to SNRs. 
    
    We assume $\gamma_{\max} = \infty$ in our calculations.
    Since the energy density distribution is a steep power-law, 
    this assumption does not significantly affect our analysis.
    
    To determine $C_\gamma$ and $\gamma_{\min}$, we adopt the common
    formalism of assuming that the energy density in the power-law
    electron distribution, $u_e$, is a time-independent fraction of the 
    shock energy density
    \begin{eqnarray}
        u_e = \epsilon_e \rho_\csm v_s^2,
    \end{eqnarray}
    where $\rho_\csm$ is the pre-shock CSM mass density.
    Clearly, $\epsilon_e < 1$, but its value is not entirely certain. 
    Studies of gamma-ray afterglows, for example, find $\epsilon_e \sim 0.1$ 
    \citep{PanaitescuKumar02,Yost+03,Chevalier+04,Marongiu+22}.
    Simulations of particle acceleration in shocks also find that about $10\%$
    of the postshock energy is in relativistic ions \citep{CaprioliSpitkovsky14}, 
    so, if electrons and ions equilibrate, then $\epsilon_e \sim 0.1$;
    however, in low-density winds it is unlikely that they
    do equilibrate, which would suggest $\epsilon_e \ll 0.1$.
    A recent study of supernova remnants finds a wide range of 
    $\epsilon_e\sim 10^{-4}-0.08$ \citep{Reynolds+21}.     
    We take $\epsilon_e=0.1$ as our fiducial value.

    Using the $\epsilon_e$ formalism we then have
    \begin{eqnarray}
        n_{e,\mathrm{nt}} = \int_{\gamma_{\min}}^\infty C_\gamma \gamma^{-p} d\gamma, \\
        \epsilon_e \rho_\csm v_s^2 = m_e c^2 \int_{\gamma_{\min}}^\infty C_\gamma \gamma^{-p+1} d\gamma 
    \end{eqnarray}
    thus,
    \begin{eqnarray}
        C_\gamma = (p-1) n_{e,\mathrm{nt}} \gamma_{\min}^{p-1} \\
        \gamma_{\min} = \frac{p-2}{p-1} \frac{\epsilon_e \rho_\csm v_s^2}{m_e c^2 n_{e,\mathrm{nt}}} ~.
    \end{eqnarray}
    If a fraction $f_\mathrm{nt}$ of the total electrons in the shocked region 
    are in the power-law component, then
    $n_{e,\mathrm{nt}} = \eta f_\mathrm{nt} \rho_\csm / (\mu_e m_p)$, 
    where $\mu_e = \rho/(n_e m_p)$ and $\eta$ is the mass compression ratio.
    Thus,
    \begin{eqnarray}
        \gamma_{\min} &=& 2.04 \frac{p-2}{p-1}
        \left(\frac{\mu_e\epsilon_e}{\eta f_\mathrm{nt}}\right) 
        v_{s,9}^2 ~,
    \end{eqnarray}
    which can be evaluated as a function of shock radius or time since
    explosion via Equations \ref{eqn:v_s_Rfunc} \& \ref{eqn:v_s_tfunc}.
    We assume $f_\mathrm{nt}=1$ to begin with, but in the 
    case that this results in $\gamma_{\min}<1$, we 
    decrease $f_\mathrm{nt}$ such that $\gamma_{\min}=1$.
    Thus, as noted in \citet{ChevalierFransson06}, we are 
    accounting for the fact that a result of $\gamma_{\min}<1$
    indicates a substantial population of thermal electrons.    
    
    In the limit that $\gamma_{\min}=1$, 
    \begin{eqnarray}
        C_\gamma = (2.99\times10^{5}~\mathrm{cm^{-3}}) (p-1) \frac{\eta f_\mathrm{nt}}{\mu_e} 
        q_\star R_{f,15}^{-2} ~,
    \end{eqnarray}
    but for $\gamma_{\min}>1$ and $p=3$,
    \begin{eqnarray}
        C_\gamma = (4.75\times10^8 \mathrm{cm^{-3}}) \frac{\mu_e \epsilon_e^2}{\eta f_\mathrm{nt}} M_c^{-10/7} E_{51}^{2} q_*^{3/7} R_{f,15}^{-18/7} ~.
    \end{eqnarray}

        \subsubsection{Cooling of Relativistic Electrons}\label{sssec:electron_cooling}

        Synchrotron emission will be diminished if relativistic electrons
        in the shocked ejecta cool. 
        A relativistic electron of energy $\sim\gamma m_e c^2$ will cool by the 
        synchrotron process in a time
        \begin{equation}
            t_\mathrm{cool} \sim \frac{\gamma m_e c^2}{\frac{4}{3} \sigma_T c u_B \gamma^2}
            \label{eqn:tcool}
        \end{equation}
        where $u_B = \epsilon_B \rho_\csm v_s^2$ can be calculated from Equation~\ref{eqn:rampress_t}.
        
        The formalism of assuming the magnetic field energy density $u_B = B^2/8\pi$ is 
        a constant fraction $\epsilon_B$ of the internal energy density is in common use
        for astrophysical shocks; yet, there is significant debate surrounding the value of
        $\epsilon_B$.
        Analyses of GRB afterglows find a wide range of values, $\epsilon_B=7.4\times10^{-4}-0.2$
        \citep[e.g.,][]{Chevalier+04,PanaitescuKumar02,Yost+03}. 
        Studies of core-collapse supernovae interacting with the progenitor wind 
        also find a wide range -- for example,
        $\epsilon_B \sim 10^{-3} q_\star^{-1}$ derived for the Type Ic SN~2002ap \citep{BjornssonFransson04}, 
        $\epsilon_B \sim 3\times10^{-4}$ for the Type IIb SN~2011dh \citep{Horesh+13}, 
        or $\epsilon_B = 0.24 (0.003)$ for the Type II SN~1987A (SN 1979C) \citep{Chevalier98}.
        Taking a theoretical approach, 
        \citet{DuffellKasen17} posit that magnetic fields should be amplified by the turbulent
        cascade of the Rayleigh-Taylor instability that occurs naturally in 
        astrophysical shocks, and derive a lower bound of $\epsilon_B=3\times10^{-3}$
        from this mechanism. 
        In this work, we take $\epsilon_B=0.1$ to be our fiducial (or ``default'') value, 
        because this is typical value assumed in SN~Ia literature; 
        but when we perform our analysis we consider a range 
        $\epsilon_B=3\times10^{-3} - 0.3$, i.e., ranging from the turbulence limit to
        essentially the maximum possible value from equipartition.
        
        After assuming a value of $\epsilon_B$, we then have
        \begin{equation}
            \frac{t_\mathrm{cool}}{t} \sim 695 \gamma^{-1} \left(\frac{\epsilon_B}{0.1}\right)^{-1} q_\star^{-1} t_{10} ~.
        \end{equation}
        Alternatively, we can use the cyclotron frequency, which for the wind 
        interaction models is
        \begin{equation}
            \nu_\mathrm{cyc} = \sqrt{\frac{2}{\pi}} \frac{q_e}{m_e c} \sqrt{u_B} 
            = (3.18~\mathrm{MHz}) t_{10}^{-1} \left(\frac{\epsilon_B}{0.1} q_\star \right)^{1/2}  ~,
        \end{equation}
        to substitute 
        \begin{equation}
            \gamma = \sqrt{\nu/\nu_\mathrm{cyc}}
                = 43.45 \sqrt{\nu_6 t_{10}} \left(q_\star \frac{\epsilon_B}{0.1}\right)^{1/4} ,
        \end{equation}
        where $\nu_6 = \nu/(6~\mathrm{GHz})$,
        and have
        \begin{equation}
            \frac{t_\mathrm{cool}}{t} \sim 16 \left(\frac{t_{10}}{\nu_6}\right)^{1/2} \left(q_\star \frac{\epsilon_B}{0.1} \right)^{-3/4}
        \end{equation}        
        In Figure~\ref{fig:selfsim_applicability} we show the limit of what may be called
        rapid cooling, where $t_\mathrm{cool}/t = 1\%$. 

        This equation is useful for anticipating that
        the compact shell CSM will 
        suffer from very rapid cooling of the relativistic electron population, 
        and relativistic electrons would only be present immediately behind the
        shock front.
        Already, we expected that detecting radio emission during interaction would
        be impossible because interaction ends within hours post-explosion and
        pre-shock CSM would absorb the radio emission -- the only hope was
        catching a radio ``tail'' from the expanding post-shock shell.
        Given the rapid cooling time for relativistic electrons, 
        we now see that a radio detection also hinges on the post-interaction
        expansion being rapid enough to slow down 
        relativistic electron cooling.
        This is our motivation for performing a SNEC simulation with the 
        same setup as \citet{PiroMorozova16}, but with finer time resolution of the
        output hydrodynamic grid focused on the earliest phases of the SN evolution
        that allows an exploration of how the gas evolves during and immediately
        after interaction with the compact CSM shell.

        We can calculate the cooling time just behind the shock front 
        in the SNEC simulation via Equation~\ref{eqn:tcool}.
        We focus on the immediate post-shock region because electrons that 
        cool immediately are not expected to be re-accelerated downstream, so 
        the immediate post-shock region contains the electrons of interest.
        The hydrodynamic quantities just behind the shock are obtained as follows.
        While $R_f \leq R_\mathrm{out}$, SNEC tracks the shock properties (zone index,
        radius, speed) over time with finer time resolution than the output 
        snapshots of the entire domain, and provides the shock solution as output.
        We use a cubic 
        interpolation (via \texttt{scipy.interpolate.interp1d})
        of the shock zone index over time to calculate 
        the zone index of the shock at the snapshot times (rounding down), 
        and use the cell-centered values of 
        hydrodynamic variables from the preceding cell to represent conditions 
        just behind the shock.
        We calculate the hydrodynamic time using the radius and speed
        at the center of the cell just behind the shock; though we note that 
        $t_\mathrm{hydro} \sim t$.
        For the cooling time calculation, 
        we have the specific internal energy and the mass density ($\rho$) in the cell,
        whose product is the internal energy density, $u_\mathrm{int}$.
        While the shock is in the CSM, $u_\mathrm{int} = u_e/\epsilon_e = u_B/\epsilon_B$.
        Since the cooling time depends on $\epsilon_B$, we explore both
        $\epsilon_B=0.003$ and $0.1$.
        If relativistic electrons do not cool rapidly in the immediate post-shock
        region, then these $\epsilon_B$ values also represent the possible 
        change in $\epsilon_B$ downstream from the shock \citep{Crumley+19}. 
        For times after $R_f = R_\mathrm{out}$, we use the properties of 
        the second-to-last zone of the domain to calculate the cooling time
        of the outermost CSM as it rarefies.
        After the shock has swept over the CSM, we assume 
        the number of (isotropic) magnetic field lines is conserved 
        in the expanding gas such that $B\propto r^{-2}$, 
        i.e., $u_B \propto r^{-4}$.

    \subsection{Synchrotron Light Curve}\label{ssec:model_lc}
    
        Our method for generating synchrotron radio light curves 
        is as in \citet{Chomiuk+16}, which is a restatement of the
        equations from \citet{Chevalier98}.
        Since we consider higher wind densities than in \citet{Chomiuk+16},
        we furthermore include the free-free absorption equations from
        \citet{Chevalier82b}.
        We note that a similar procedure could be used to approximate
        a light curve from the SNEC compact shell simulation, but is not
        performed in this work as we find that relativistic electrons do not 
        survive ($C_\gamma=0$), as we will show in \S\ref{ssec:shell_results}.
        
        The flux density
        assuming synchrotron self-absorption (SSA) only and $p=3$
        is
        \begin{equation}
            F_\nu = 6.67\times10^{-30} R_f^2 D_L^{-2} B^{-1/2} \nu^{5/2} 
            \left[1 - e^{-\left(\nu/\nu_1\right)^{7/2}}\right] ~,
            \label{eqn:ssa_flux}
        \end{equation}
        where $D_L$ is the distance to the target, $\nu$ is frequency, 
        $B$ is the magnetic field strength, and all quantities are in cgs units
        such that the final units of $F_\nu$ are $\mathrm{erg~s^{-1}~cm^{-2}~Hz^{-2}}$.
        The quantity $\nu_1$ is the frequency at which the SSA optical depth
        is unity, and is given by
        \begin{equation}
            \nu_1 = (47.6~\mathrm{MHz}) R_f^{2/7} f^{2/7} C_\gamma^{2/7} B^{5/7} ~,
        \end{equation}
        where $f$ is the filling factor of the emitting region, defined 
        via $V_\mathrm{em} = f (4 \pi R_f^3/3)$.
        Then, for $n=10$ and $s=2$, we use
        $f = 1 - (R_r/R_f)^3 = 0.5$ \citep{Chevalier82a}.

        Since the CSM densities considered in this work span a wide 
        range and the observations cover very early times, 
        we will consider the effects of free-free absorption by
        the unshocked CSM in addition to the 
        synchrotron self-absorption from the shocked CSM.
        We use Equation 22 of \citet{Chevalier82b}, to compute 
        the free-free optical depth in the case of an
        infinite-extent wind as
        \begin{equation}
            \tau_\mathrm{ff} = 4.35~q_\star^2 R_{f,15}^{-3}
                              \left(\frac{\nu}{6~\mathrm{GHz}}\right)^{-2}
        \end{equation}
        where we have replaced the maximum material velocity with $R_r/t$.
        We account for external absorption by multiplying 
        the flux from Equation \ref{eqn:ssa_flux} by $\exp(-\tau_\mathrm{ff})$. 
        Due to the early times of our observations, free-free absorption is 
        dominant even for low mass-loss rates.

\section{Results}\label{sec:results}

In this section, we use the calculations described 
in \S~\ref{sec:models} to analyze our early-time radio dataset
and draw conclusions about the progenitor systems for these SNe~Ia.

    \subsection{Wind-Like CSM}\label{ssec:wind_results}
    
\begin{figure}
            \centering
            \includegraphics[width=\linewidth]{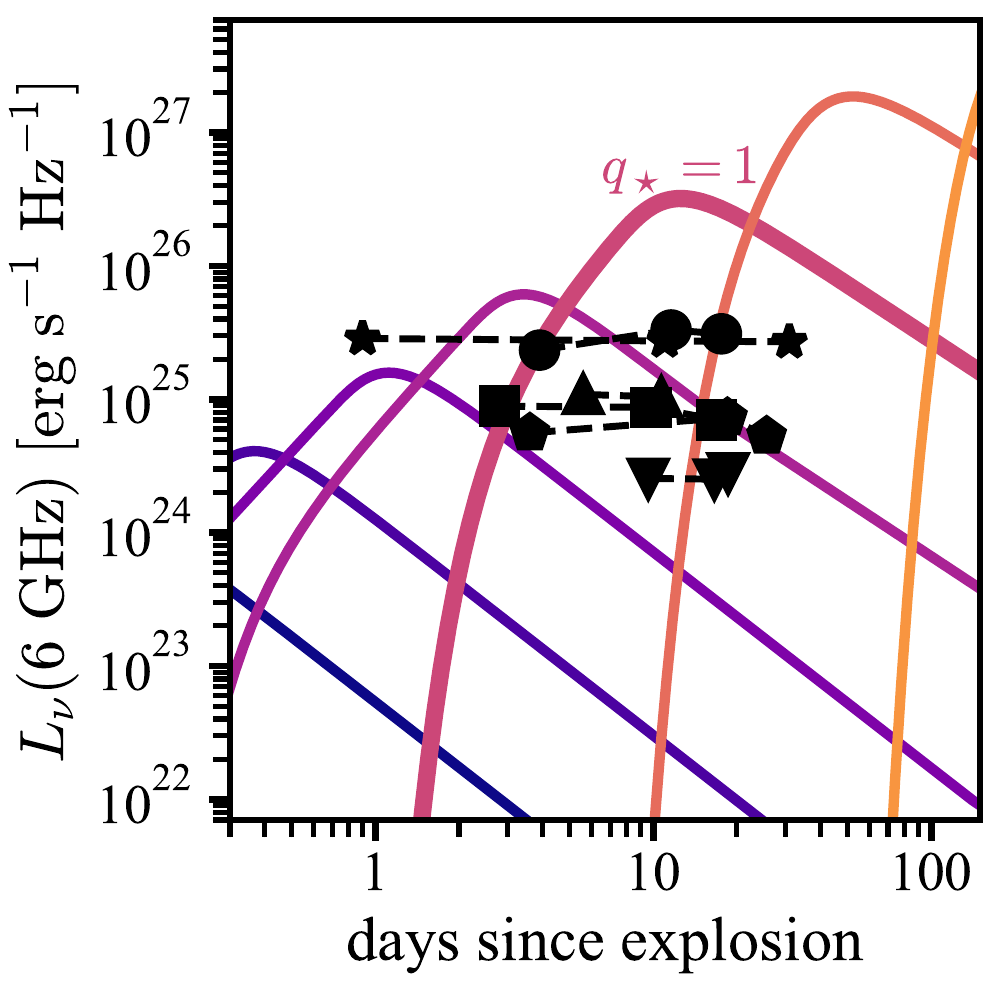}
            \caption{Wind interaction 6 GHz synchrotron light curves, 
            assuming fiducial parameters, in decade increments of
            $q_\star$ from $10^{-4}$ (blue) to $10^2$ (yellow-orange) -- 
            i.e., $\dot{M}/v_w = 10^{-12} - 10^{-6} ~\mathrm{(M_\odot~yr^{-1})/(km~s^{-1})}$.
            Markers connected by dashed lines show our radio observations
            of SN~2019np (squares), SN~2019ein (circles), SN~2020rcq (down-triangles), 
            SN~2020uxz (stars), SN~2021qvv (pentagons), and SN~2021smj (up-triangles).
            }
            \label{fig:light_curves}
        \end{figure}

        Figure~\ref{fig:light_curves} shows the synchrotron light curves 
        that result from interaction with a wind with density 
        $10^{-4} \leq q_\star \leq 10^{2}$
        compared to the limits obtained for our six targets. 
        We see that winds with $q_\star < 10^{-2}$ peak within a day post-explosion
        and therefore escape detection even from the earliest observations. 
        At higher densities than this, $10^{-2} < q_\star < 1$, 
        we see a power-law rise characteristic of the dominance of 
        synchrotron self-absorption. 
        In the $q_\star = 0.1, 1$ cases, we see that this power-law rise is 
        preceded by an exponential rise, and the two highest density cases
        have exponential rises; this is the signature of diminishing 
        free-free absorption.
        From this figure it is easy to see that 
        our early-time observations will have the most constraining power
        for $10^{-2} < q_\star < 10$ winds, which includes most of the 
        CSM wind scenarios we are investigating as shown in Figure \ref{fig:selfsim_applicability}. 
        
\begin{figure}
            \centering
            \includegraphics[width=\linewidth]{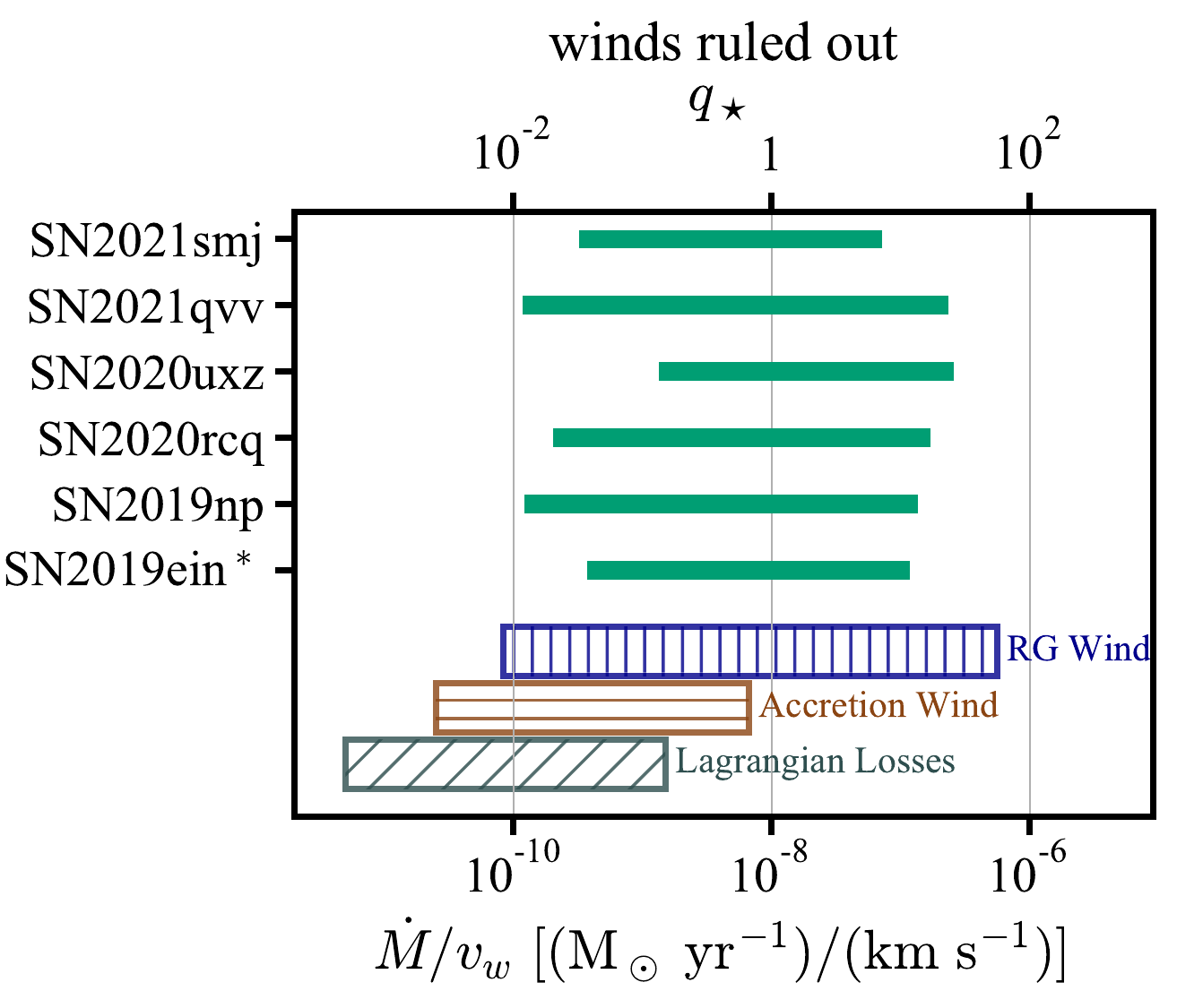}
            \caption{Solid green bars show the span of winds ruled out by radio observations,
            assuming fiducial shock parameters -- lower and 
            higher density winds are allowed.
            Hatched regions show expectations for wind densities from 
            the physical mechanisms discussed in the text.
            }
            \label{fig:wind_constraint}
        \end{figure}

        Using a grid of light curves varying $q_\star$, 
        we are able to determine which winds our radio limits could
        detect.
        The span of wind densities thus ruled out is shown in 
        Figure~\ref{fig:wind_constraint}.
        The low-density edge to these limits comes from the early-time
        and low-luminosity peaks of low-$q_\star$ light curves.
        The high-density edge is due to the late-time peak of high-$q_\star$
        light curves that results from free-free absorption. 
        
        We see that for most SNe, we are able to rule out winds down to 
        densities $\dot{M}/v_w \sim 10^{-9} - 10^{-10}~\mathrm{(M_\odot~yr^{-1})/(km~s^{-1})}$.
        This represents a ruling out of the lowest density symbiotic winds, 
        and it also rules out a significant span of the possible winds associated 
        with high accretion rates.

\begin{figure}
            \centering
            \includegraphics[width=\linewidth]{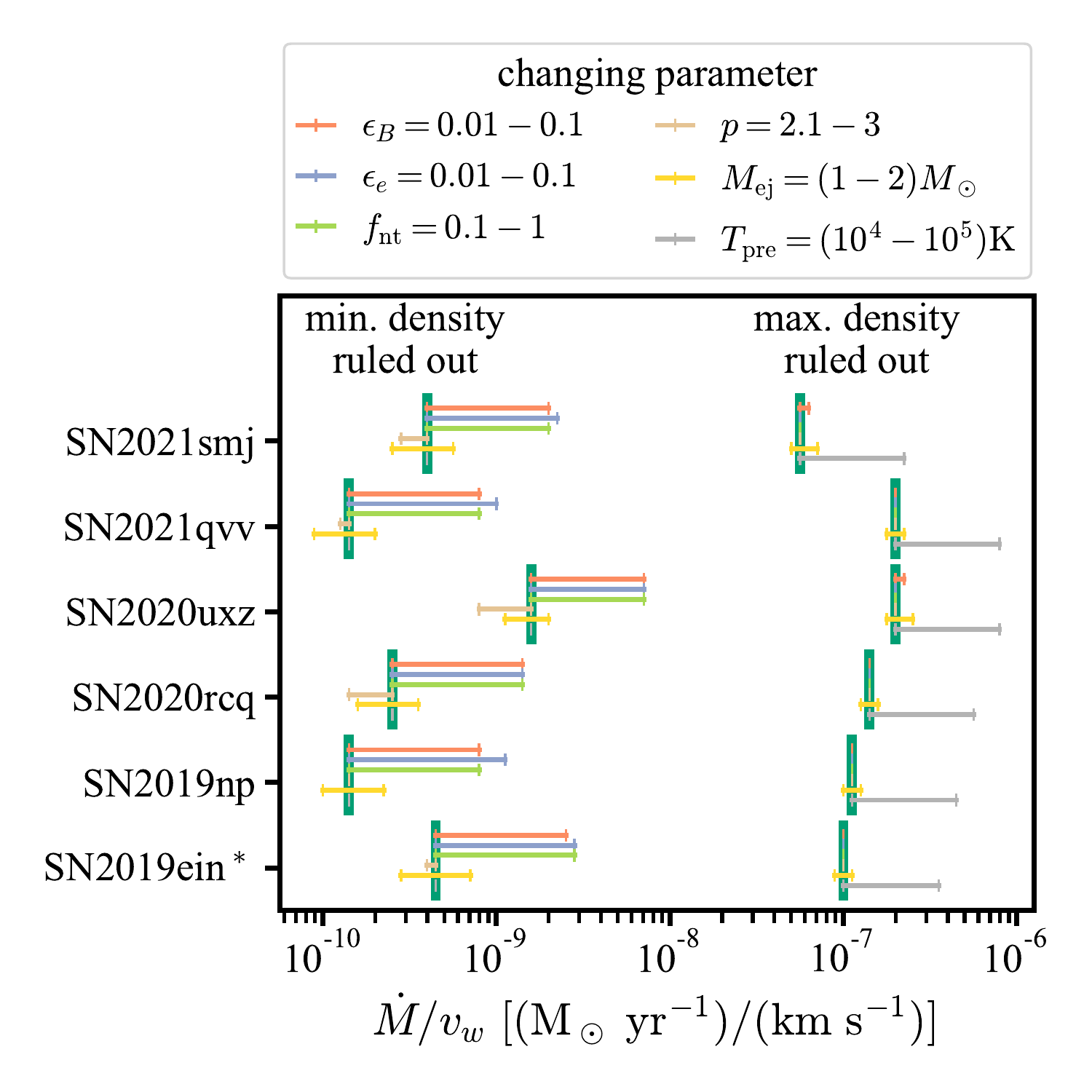}
            \caption{The effect of model parameter assumptions on our 
            results for the wind model. 
            Vertical green bars indicate the minimum and maximum 
            wind density ruled out under our ``fiducial'' assumptions
            (see Figure~\ref{fig:wind_constraint}). 
            Horizontal error bars on these show the effect of changing the 
            magnetic field energy density ($\epsilon_B$, red), 
            nonrelativistic electron energy density ($\epsilon_e$, blue),  
            fraction of electrons in the power-law population ($f_\mathrm{nt}$, green),
            power-law index of the relativistic electrons ($p$, brown), 
            supernova ejecta mass ($M_\mathrm{ej}$, yellow), and 
            temperature of the preshock CSM ($T_\mathrm{pre}$, grey). 
            See text for description of the effect of distance uncertainty.
            }
            \label{fig:mod_err}
        \end{figure}

        However, as discussed in \S~\ref{sec:models}, 
        these constraints are subject to underlying uncertainties.
        Our default values for various model parameters
        (e.g., $\epsilon_B$) were chosen to reflect values that are 
        commonly used in the literature of SNe interacting with winds.
        In Figure~\ref{fig:mod_err} we show how changing 
        each parameter affects the constraint for each of our targets.
        A decade (``1 dex'') decrease in $\epsilon_B$, $\epsilon_e$, 
        or $f_\mathrm{nt}$ results in a $\sim0.8$ dex increase on 
        the minimum wind density we can rule out, weakening the 
        constraining power of the observations.
        On the other end, we see that a 1 dex increase in the 
        absorbing (pre-shock) CSM temperature, $T_\mathrm{pre}$, 
        increases the window of CSM ruled out by $\sim0.5$ dex.
        Smaller effects are seen when changing the 
        power-law slope of the non-thermal electrons ($p$) 
        or the ejecta mass, since these parameters do not suffer
        orders of magnitude uncertainty in their values.
        We see that if $p=2.1$, our observations can 
        rule out slightly lower density winds in some cases.
        For ejecta mass, we consider decreasing from $1.38~M_\odot$ to 
        $1~M_\odot$ (sub-Chandrasekhar explosion, \cite{}) as well as 
        increasing to $2~M_\odot$ (super-Chandrasekhar explosion), 
        and we see that this has a small impact on the lowest density wind
        that we can constrain. 
        
        Furthermore, we investigated the effect of the assumed distance on 
        our wind density limits, by re-analyzing the radio data with 
        redshift-independent distances from the NASA/IPAC Extragalactic 
        Database for 
        SNe 2019np \citep{Willick+97,Theureau+07,Sorce+14}, 
        2020uxz \citep{Tully+09,Tully+13,Tully+16,Springob+09},
        and 2021qvv \citep{Jordan+05,Jordan+07,Villegas+10}
        and the error range on the distance to SN~2019ein given in <
        \citet{Pellegrino2020}. 
        For SNe 2019ein and 2021qvv, the distance differences
        are $<0.1~\mathrm{dex}$, whereas the distance varies more
        for SN~2019np and SN~2020uxz ($\lesssim0.3~\mathrm{dex}$
        and $\lesssim0.15~\mathrm{dex}$, respectively). 
        The resulting variation in the minimum density 
        constrained is of a similar magnitude.
        Therefore, we conclude that distance uncertainty is 
        subdominant to the uncertainties from underlying physics.
        
        For the purposes of ruling out low density winds around SNe~Ia, 
        it is clear that the uncertainties in $\epsilon_B$, $\epsilon_e$, 
        and $f_\mathrm{nt}$ are key issues.
        Considering that the true values could be even lower than what we 
        chose for our illustration, and that their effects compound, 
        it is easy to see a pessimistic combination that renders the radio 
        observations totally unconstraining.
        As sensitive radio data are already in hand from this and other
        studies, we therefore emphasize the great importance of research, 
        especially theoretical work, on 
        magnetic field amplification and electron acceleration in 
        SN shocks.

        \subsubsection{Higher-Density Winds Disfavored by Optical Data}\label{sssec:high_dens_wind}
        
        Looking at the $g$- and $r$-band optical light curves from the 
        Zwicky Transient Facility via the 
        ALeRCE ZTF Explorer\footnote{https://alerce.online/}
        and the optical spectra on the Transient Name Server\footnote{https://www.wis-tns.org/},
        we see no obvious indication of interaction with high-density material
        for our six targets, which can show up in several ways. 
        Considering extended winds, we can use this to complement the
        radio limits on wind density on the high-$q_\star$ side.
        From \citet{Ofek+13}, the fact that CSM would affect the 
        rise time limits the mass loss
        rate to $<2.7\times10^{-4}~\mathrm{(M_\odot~yr^{-1})/(km~s^{-1})}$.
        Narrow H$\alpha$ line formation from recombination is created at the 
        level of $\sim5\times10^{39}~\mathrm{erg~s^{-1}}$ 
        for $\dot{M}/v_w = 10^{-5}~\mathrm{(M_\odot~yr^{-1})/(km~s^{-1})}$,
        which is approximately where free-free absorption cuts off the 
        utility of the early-time radio limits.
        Densities higher than this would have a more dramatic effect -- 
        \citet{Silverman+13} estimate that SNe~Ia-CSM have
        $\dot{M}/v_w = 10^{-3}~\mathrm{(M_\odot~yr^{-1})/(km~s^{-1})}$,
        and in those cases there is a very strong continuum in the
        optical spectra in addition to the narrow and broad
        H$\alpha$ emission features.
        There is no obvious indication of any of these features in the 
        optical data, disfavoring the possibility that there is 
        CSM more dense than the range we are able to rule out with our
        radio observations.
        However, a more in-depth analysis of the 
        optical data would be necessary for a definitive conclusion --
        for example, only after constructing a bolometric light curve do
        \citet{Sai+22} see a low-level (5\%) excess in one of our targets, 
        SN~2019np. 
        In the case of SN~2019np, \citet{Sai+22} present spectra at very early times
        that show no evidence of CSM interaction emission, and their analysis
        indicates that this small excess is caused by an extended $^{56}$Ni
        distribution in the ejecta.

        \subsubsection{Prior Limits on Wind Scenarios}\label{sssec:lit_wind}

        \citet{Chomiuk+16} and \citet{Lundqvist+20} present the collected radio
        observations of SNe~Ia from the literature (see references therein), and
        \citet{Hosseinzadeh+22} provides an additional case with SN~2021aefx.
        SN~2021aefx is an early-excess SN Ia, but the radio observations at
        $t\gtrsim4~\mathrm{days}$ post-explosion probe an extended medium rather than
        possible compact CSM so we interpret these observations in a wind-like
        CSM context, as do \citet{Hosseinzadeh+22} -- see also \S~\ref{ssec:shell_results}.
        We apply the same methodology to these collected datasets as to our new 
        observations to discuss the impact of our six new SNe~Ia to our overall 
        understanding of SN~Ia progenitors. 
        Here we confine ourselves to the normal, 
        91T-like (``shallow Silicon''), 
        and 91bg-like (``cool'') SNe~Ia from \citet{Chomiuk+16}.
        
\begin{figure}
            \centering
            \includegraphics[width=\linewidth]{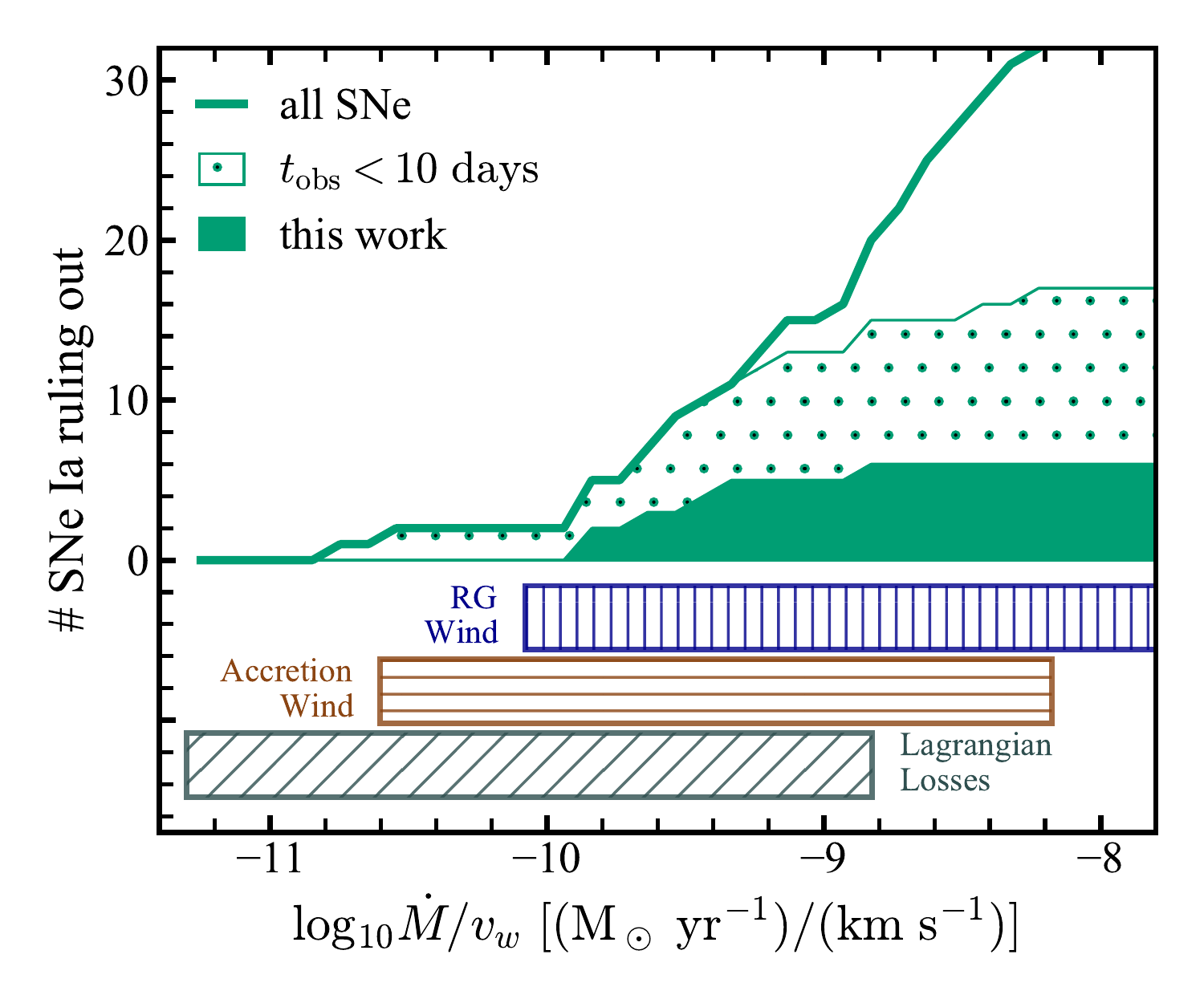}
            \caption{A cumulative histogram of how many SNe~Ia can rule out low-density winds,
            comparing this paper (green filled) to
            published SNe~Ia with radio observations within 10 days (green hatched) 
            and the full sample of all published SN~Ia radio observations (green solid line).
            Our observations represent a significant increase to the number of 
            SNe~Ia that can rule out winds with $q_\star < 0.1$. 
            }
            \label{fig:all_obs_histogram}
        \end{figure}

        Figure~\ref{fig:all_obs_histogram} shows the result of this analysis
        as a histogram of how many SNe~Ia are known with radio data that can 
        constrain winds of different densities, with a focus on the lowest densities. 
        We indicate how many of the SNe~Ia at each density come from our sample,
        the literature sample of SNe~Ia with 
        observations within 10 days post-explosion,
        and the full literature sample (including SNe~Ia first observed later
        than 10 days post-explosion) 
        for an analysis with our fiducial parameter set.
        For the full sample, we also indicate how the histogram may shift
        with different choices of parameters, as in Figure~\ref{fig:wind_constraint}. 
        Note that the two events at the \textit{very} lowest densities
        are SN 2011fe and SN 2014J 
        \citep{Chomiuk2012,PerezTorres+14}.    
    
        We show by this plot that the six SNe~Ia presented in this
        manuscript represent a significant increase in the number of SNe~Ia 
        with constraints at these low wind densities.
        Again, this is because interaction with low-density winds 
        creates a radio light curve that peaks at early times 
        (see Figure~\ref{fig:light_curves}).
        The relative impact of our observations increases as one 
        looks to lower density constraints -- the sample nearly doubles
        events that can constrain the accretion wind and Lagrangian losses 
        scenarios. 
        The predominance of SNe~Ia with early-time observations 
        at the lowest density end of this histogram highlights the 
        necessity of rapid follow-up observations, 
        particularly if it turns out that the microphysics of the shock 
        acceleration and magnetic field amplification is unfavorable 
        (i.e., the weakest limits are correct).

    \subsection{Compact CSM Shells}\label{ssec:shell_results}
    
        Recall from \S~\ref{sssec:electron_cooling} that our hope for radio emission 
        in the case of compact shell interaction is that
        after the shock has swept over the CSM, the last, small parcel of
        gas may be accelerated so rapidly that its cooling time rapidly increases and 
        relativistic electrons can survive.
        We now assess this possibility with our SNEC simulation.

        Using the method described in \S\ref{sssec:electron_cooling}, we
        calculate cooling time for $\gamma$ corresponding to 6 GHz;
        however, we find that while the shock is in the CSM, the energy density is so high 
        that $B \gtrsim 10^3~\mathrm{G}$ and $\nu_\mathrm{cyc} \gtrsim 10~\mathrm{GHz}$
        ($\nu_\mathrm{cyc} \sim 3B~\mathrm{MHz}$).
        Physically, this means that even without cooling and with $\epsilon_B=0.003$, 
        synchrotron emission would not extend down to 6 GHz 
        until the energy density drops after shock breakout.
        In our calculation of the cooling time, we use $\gamma=1$ when 
        $\nu_\mathrm{cyc}>6~\mathrm{GHz}$.

\begin{figure}
            \centering
            \includegraphics[width=\linewidth]{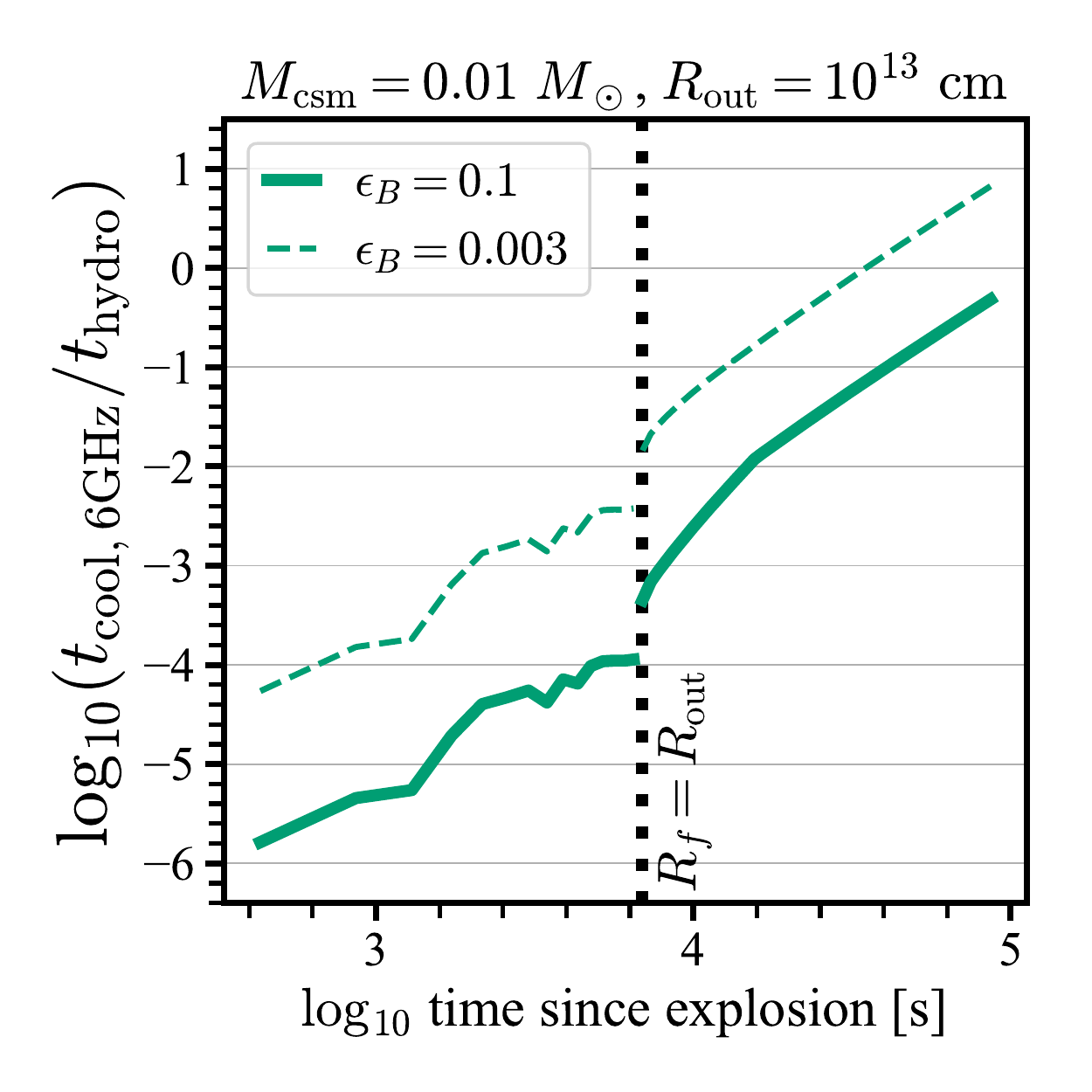}
            \caption{The cooling time of relativistic electrons that produce
            6 GHz emission compared to the hydrodynamic timescale (i.e., the efficiency
            of relativistic electron cooling) as a function of time,
            assuming $\epsilon_B=0.1$ (solid) and $\epsilon_B=0.003$ (dashed).
            $R_f=R_\mathrm{out}$ marks the time the
            shock crosses the compact CSM shell. 
            We see that cooling of the relativistic electrons is likely too efficient for
            relativistic electrons to survive despite the rapid expansion of the CSM
            after the interaction.
            }
            \label{fig:shell_cooling}
        \end{figure}

        Figure~\ref{fig:shell_cooling} shows the cooling efficiency of relativistic
        electrons over time 
        immediately behind the shock front in the SNEC simulation.
        We indicate the time that the shock front overtakes the 
        outer edge of the CSM ($R_f = R_\mathrm{out}$), and reiterate that 
        we do not expect 6 GHz emission prior to this point due to high $\nu_\mathrm{cyc}$.
        Prior to this point, we see that cooling is extremely efficient,
        although it is trending to longer cooling times.
        After the shock crosses the CSM, the cooling time does, indeed, 
        increase very rapidly. 
        However, it remains very short compared to the hydrodynamic time.
        Cooling times are longer for the lower $\epsilon_B$, because 
        the power radiated by synchrotron emission is lower and -- once the 
        shock has crossed the CSM -- $\gamma$ is also lower.
        Yet even in the lower $\epsilon_B$ case, the situation looks grim for
        relativistic electron survival. 
    
        For a variety of reasons, we determine that interaction with  
        compact shells is unlikely
        to produce detectable radio emission from the 
        synchrotron process.
        Prior to the shock crossing the CSM, free-free absorption will 
        stifle radio emission and cooling will severely limit the volume of
        relativistic electrons; furthermore, we have seen that the magnetic
        field strength may be so high that synchrotron emission is pushed to
        frequencies above the radio range.
        Once the shock crosses the CSM, our hope was that the rapid rarification 
        of the outermost CSM, and thus increase of cooling time, 
        might preserve a small fraction of relativistic electrons.
        However, it seems from our estimates that the cooling time does not
        increase fast enough to preserve the relativistic population,
        though the issue with the magnetic field strength is resolved such that
        there could be synchrotron emission at radio frequencies.

        It may be worth determining in future work whether the X-rays generated from
        cooling relativistic electrons can at any point escape the envelope,
        producing a complementary X-ray/gamma-ray flash to the optical/UV excess, analogous to the shock breakout from
        core-collapse SNe \citep{NakarSari10}, which could potentially leverage the rapid follow-up capabilities of a space telescope like the \textit{Neil Gehrels Swift Observatory}. 
        A thermal X-ray signal is not expected from our models because
        the shock-heated gas, equilibrating with the radiation field, 
        reaches only $kT\sim0.01~\mathrm{keV}$. 
        We mention this to contrast with the case of the calcium-strong
        transient SN~2019ehk -- 
        \citet{Jacobson-Galan+20} interpret the
        \textit{Swift} X-ray detection of
        SN~2019ehk in the context of interaction with a compact shell,
        but attribute the X-ray emission to thermal emission from 
        $kT>10~\mathrm{keV}$ gas. 
        The emission from cooling high-energy electrons would be
        non-thermal and of much shorter duration.

        Finally, we note that the radio prospects 
        may not be not entirely hopeless, despite our simulation result. 
        In the first place, we have not performed a detailed particle-in-cell
        simulation that tracks the evolution of the ion and electron populations 
        as they interact with each other and the magnetic field. 
        Such a simulation would need to also include the effects of 
        radiative cooling and photon trapping in this dense environment, 
        and therefore represents a significant research project outside the scope
        of the current work.
        However, even if such a calculation confirmed our own, there  
        are some remaining potential avenues for radio emission in
        the compact shell scenario:
        where the CSM is aspherical, perhaps the ejecta interact with significantly
        lower-density CSM in some fraction of the volume, or, 
        if there is a low-density wind outside the compact shell, 
        radio emission could arise from the interaction with this wind.
        Indeed, for the case of SN~2021aefx, the analyses of both \citet{Hosseinzadeh+22}
        and this work are focused on constraining the presence of an extended medium
        outside a possible compact shell.
        Further work predicting the environment out to $r\sim10^{16}~\mathrm{cm}$
        and assessing how initial interaction with a dense shell would change the density and
        velocity profile of the fast ejecta as it propagates into an extended medium
        are needed to assess these possibilities.

\section{Summary}\label{sec:conc}

In this paper, we have presented Jansky VLA 6 GHz observations
within 10 days post-explosion for six SNe~Ia with the 
goal of constraining Type Ia supernova progenitors via
their circumstellar environments.

The nature of the mass-donor companion that ignites CO~WDs 
as SNe Ia remains a mystery, despite decades of effort. 
Whether mass donation happens via accretion or merger, 
whether explosion occurs at the center or surface, 
whether the companion is degenerate, 
are all matters of debate -- and each possibility may
be at play to create a tapestry of progenitor scenarios 
whose relative contributions are to be accounted. 

Characterizing the circumstellar environments of SNe~Ia
can help to shed light on these questions because
different progenitor scenarios create different environments
in the millennia leading up to explosion. 
In this paper, we were particularly interested in low-density winds
from accretion scenarios that would leave no optical signature and 
the compact shells of WD merger scenarios that may create
observed early-time optical excesses in SN~Ia light curves.
In the first case, radio observations are one of the only  ways to 
see the CSM; in the second case, a radio signal would be
a clear tie to non-thermal electron populations characteristic of
shocks to clarify the origin of the emission. 

Our observations can constrain the lowest density
winds from red giant companions, optically thick accretion winds,
and the loss of material from the outer Lagrangian point of the
disk (\S~\ref{ssec:wind_hydro}, \ref{ssec:wind_results}). 
To analyze the observations in this context, we employ the 
self-similar model of \citet{Chevalier82a}, which works well for 
extended winds (Figures~\ref{fig:selfsim_applicability} \& \ref{fig:light_curves}).
With standard (fiducial) assumptions about synchrotron emission, 
the observations rule out winds down to
$(10^{-9} - 10^{-10})~\mathrm{M_\odot~yr^{-1}/(km~s^{-1})}$,
depending on the target, which represents
the lowest density red giant winds and a large portion of 
possible accretion winds (Figure~\ref{fig:wind_constraint}).
In our analysis we have accounted for uncertainties in the microphysics of
magnetic field amplification and electron energy distribution that
affect the low-density limit (Figure~\ref{fig:mod_err}).
Densities above $\sim 10^{-7}~\mathrm{M_\odot~yr^{-1}/(km~s^{-1})}$
are not ruled out by the radio observations of these targets, 
but are disfavored by the 
lack of optical signatures of interaction.
These six events represent a substantial increase in the number of 
SNe~Ia with radio observations with such low density limits (Figure~\ref{fig:all_obs_histogram}).

Toward the aim of understanding the origin of early-time excess in the 
optical/UV light curves of some SNe~Ia, we have assessed the possibility
of radio observations detecting shock emission from the interaction of
SN~Ia ejecta with a dense CSM truncated at $R_\mathrm{out}=10^{13}~\mathrm{cm}$
(\S~\ref{ssec:shell_hydro}, \ref{ssec:shell_results}). 
CSM masses between $(0.01-0.1)~M_\odot$ have been suggested for 
creating the optical/UV excesses, and 
through analytic calculations we have shown that such masses
require radiation hydrodynamics simulations to calculate the evolution 
of the shock wave and that radio emission will be severely affected 
by the cooling of relativistic electrons in the dense gas
(Figure~\ref{fig:selfsim_applicability}, \S~\ref{sec:models}). 
To assess the possibility that the rapid rarefication of the CSM \textit{after} the
shock has swept over the CSM could dramatically increase the cooling time
of electrons and allow some to survive and produce radio emission, 
we have performed radiation hydrodynamic calculations with SNEC.
Our model of a $0.01~M_\odot$ shell extending to 
$10^{13}~\mathrm{cm}$ is an expansion of the model suite presented in 
\citet{PiroMorozova16} to a larger radius and lower CSM mass. 
This model has the same CSM as in the simulation by \citet{Jiang+21}
to reproduce the early excess of SN~2020hvf, but we assume typical SN~Ia 
explosion parameters whereas their simulation had higher-mass ejecta than normal.
We find that the relativistic electron cooling time remains very short 
even through the rarefaction phase, and we therefore do not expect 
the compact shell interaction scenario to produce radio emission
at any phase.
There may, perhaps, be an X-ray flare from the cooling relativistic electrons
at the very end of interaction. 
If the WD merger or accretion from a WD scenarios produce a more extended
and less dense circumstellar environment, it may be possible for radio emission 
to be produced from interaction with this more extended medium.

Overall, we see that our deep, early-time observations serve to 
probe the existence of low-density winds around SNe~Ia, 
even accounting for uncertainties in the underlying physics 
of magnetic field amplification. 
With these observations we significantly increase the sample of 
observations that can constrain not only red giant winds, but the
lower density accretion winds from high accretion rates in Roche lobe 
overflow.
Detailed tracking of the shock evolution through a 
high density, compact CSM reveals that relativistic electrons cannot 
survive to produce synchrotron emission.
This is unfortunate, because such emission would unambiguously 
indicate the presence of a shock in SNe~Ia with short-duration
optical/UV excess to determine the physical origin of this emission.

%
%
\acknowledgments

We thank Ellen Zweibel for useful discussion in the 
preparation of this manuscript and the anonymous referee for their comments.
This research was supported by NSF grants AST-1907790 and 2107070.
Time domain research by the University of Arizona team and D.J.S.\ is supported by NSF grants AST-1821987, 1813466, 1908972, \& 2108032, and by the Heising-Simons Foundation under grant \#2020-1864. SKS also acknowledges support from the CCAPP fellowship.
The National Radio Astronomy Observatory is a facility of the National Science Foundation operated under cooperative agreement by Associated Universities, Inc. Michigan State University occupies the ancestral, traditional and contemporary 
lands of the Anishinaabeg – Three Fires Confederacy of Ojibwe, Odawa and Potawatomi peoples. 
The university resides on land ceded in the 1819 Treaty of Saginaw.

%
%

\software{
SNEC \citep{Morozova+15}, 
SciPy \citep{SciPy}, 
NumpPy \citep{NumPy}, 
Astropy \citep{Astropy},
Matplotlib \citep{Matplotlib}
}

\bibliography{refs.bbl}

\begin{thebibliography}{}
\expandafter\ifx\csname natexlab\endcsname\relax\def\natexlab#1{#1}\fi
\providecommand{\url}[1]{\href{#1}{#1}}
\providecommand{\dodoi}[1]{doi:~\href{http://doi.org/#1}{\nolinkurl{#1}}}
\providecommand{\doeprint}[1]{\href{http://ascl.net/#1}{\nolinkurl{http://ascl.net/#1}}}
\providecommand{\doarXiv}[1]{\href{https://arxiv.org/abs/#1}{\nolinkurl{https://arxiv.org/abs/#1}}}

\bibitem[{{Abbott} {et~al.}(2019){Abbott}, {Allam}, {Andersen}, {Angus},
  {Asorey}, {Avelino}, {Avila}, {Bassett}, {Bechtol}, {Bernstein}, {Bertin},
  {Brooks}, {Brout}, {Brown}, {Burke}, {Calcino}, {Carnero Rosell}, {Carollo},
  {Carrasco Kind}, {Carretero}, {Casas}, {Castander}, {Cawthon}, {Challis},
  {Childress}, {Clocchiatti}, {Cunha}, {D'Andrea}, {da Costa}, {Davis},
  {Davis}, {De Vicente}, {DePoy}, {Desai}, {Diehl}, {Doel}, {Drlica-Wagner},
  {Eifler}, {Evrard}, {Fernandez}, {Filippenko}, {Finley}, {Flaugher}, {Foley},
  {Fosalba}, {Frieman}, {Galbany}, {Garc{\'\i}a-Bellido}, {Gaztanaga},
  {Giannantonio}, {Glazebrook}, {Goldstein}, {Gonz{\'a}lez-Gait{\'a}n},
  {Gruen}, {Gruendl}, {Gschwend}, {Gupta}, {Gutierrez}, {Hartley}, {Hinton},
  {Hollowood}, {Honscheid}, {Hoormann}, {Hoyle}, {James}, {Jeltema}, {Johnson},
  {Johnson}, {Kasai}, {Kent}, {Kessler}, {Kim}, {Kirshner}, {Kovacs}, {Krause},
  {Kron}, {Kuehn}, {Kuhlmann}, {Kuropatkin}, {Lahav}, {Lasker}, {Lewis}, {Li},
  {Lidman}, {Lima}, {Lin}, {Macaulay}, {Maia}, {Mandel}, {March}, {Marriner},
  {Marshall}, {Martini}, {Menanteau}, {Miller}, {Miquel}, {Miranda}, {Mohr},
  {Morganson}, {Muthukrishna}, {M{\"o}ller}, {Neilsen}, {Nichol}, {Nord},
  {Nugent}, {Ogando}, {Palmese}, {Pan}, {Plazas}, {Pursiainen}, {Romer},
  {Roodman}, {Rozo}, {Rykoff}, {Sako}, {Sanchez}, {Scarpine}, {Schindler},
  {Schubnell}, {Scolnic}, {Serrano}, {Sevilla-Noarbe}, {Sharp}, {Smith},
  {Soares-Santos}, {Sobreira}, {Sommer}, {Spinka}, {Suchyta}, {Sullivan},
  {Swann}, {Tarle}, {Thomas}, {Thomas}, {Troxel}, {Tucker}, {Uddin}, {Walker},
  {Wester}, {Wiseman}, {Wolf}, {Yanny}, {Zhang}, {Zhang}, \& {DES
  Collaboration}}]{DES19a}
{Abbott}, T.~M.~C., {Allam}, S., {Andersen}, P., {et~al.} 2019, \apjl, 872, L30

\bibitem[{{Ashall} {et~al.}(2022){Ashall}, {Lu}, {Shappee}, {Burns}, {Hsiao},
  {Kumar}, {Morrell}, {Phillips}, {Shahbandeh}, {Baron}, {Boutsia}, {Brown},
  {DerKacy}, {Galbany}, {Hoeflich}, {Krisciunas}, {Mazzali}, {Piro},
  {Stritzinger}, \& {Suntzeff}}]{Ashall+22}
{Ashall}, C., {Lu}, J., {Shappee}, B.~J., {et~al.} 2022, arXiv e-prints,
  arXiv:2205.00606

\bibitem[{{Astropy Collaboration} {et~al.}(2013){Astropy Collaboration},
  {Robitaille}, {Tollerud}, {Greenfield}, {Droettboom}, {Bray}, {Aldcroft},
  {Davis}, {Ginsburg}, {Price-Whelan}, {Kerzendorf}, {Conley}, {Crighton},
  {Barbary}, {Muna}, {Ferguson}, {Grollier}, {Parikh}, {Nair}, {Unther},
  {Deil}, {Woillez}, {Conseil}, {Kramer}, {Turner}, {Singer}, {Fox}, {Weaver},
  {Zabalza}, {Edwards}, {Azalee Bostroem}, {Burke}, {Casey}, {Crawford},
  {Dencheva}, {Ely}, {Jenness}, {Labrie}, {Lim}, {Pierfederici}, {Pontzen},
  {Ptak}, {Refsdal}, {Servillat}, \& {Streicher}}]{Astropy}
{Astropy Collaboration}, {Robitaille}, T.~P., {Tollerud}, E.~J., {et~al.} 2013,
  \aap, 558, A33

\bibitem[{{Bj{\"o}rnsson} \& {Fransson}(2004)}]{BjornssonFransson04}
{Bj{\"o}rnsson}, C.-I., \& {Fransson}, C. 2004, \apj, 605, 823

\bibitem[{{Bloom} {et~al.}(2012){Bloom}, {Kasen}, {Shen}, {Nugent}, {Butler},
  {Graham}, {Howell}, {Kolb}, {Holmes}, {Haswell}, {Burwitz}, {Rodriguez}, \&
  {Sullivan}}]{Bloom+12}
{Bloom}, J.~S., {Kasen}, D., {Shen}, K.~J., {et~al.} 2012, \apjl, 744, L17

\bibitem[{{Bochenek} {et~al.}(2018){Bochenek}, {Dwarkadas}, {Silverman}, {Fox},
  {Chevalier}, {Smith}, \& {Filippenko}}]{Bochenek+18}
{Bochenek}, C.~D., {Dwarkadas}, V.~V., {Silverman}, J.~M., {et~al.} 2018,
  \mnras, 473, 336

\bibitem[{{Branch}(1998)}]{Branch98}
{Branch}, D. 1998, \araa, 36, 17

\bibitem[{{Branch} {et~al.}(1995){Branch}, {Livio}, {Yungelson}, {Boffi}, \&
  {Baron}}]{Branch+95}
{Branch}, D., {Livio}, M., {Yungelson}, L.~R., {Boffi}, F.~R., \& {Baron}, E.
  1995, \pasp, 107, 1019

\bibitem[{{Cao} {et~al.}(2015){Cao}, {Kulkarni}, {Howell}, {Gal-Yam},
  {Kasliwal}, {Valenti}, {Johansson}, {Amanullah}, {Goobar}, {Sollerman},
  {Taddia}, {Horesh}, {Sagiv}, {Cenko}, {Nugent}, {Arcavi}, {Surace},
  {Wo{\'z}niak}, {Moody}, {Rebbapragada}, {Bue}, \& {Gehrels}}]{Cao+15}
{Cao}, Y., {Kulkarni}, S.~R., {Howell}, D.~A., {et~al.} 2015, \nat, 521, 328,
  \dodoi{10.1038/nature14440}

\bibitem[{{Caprioli} \& {Spitkovsky}(2014)}]{CaprioliSpitkovsky14}
{Caprioli}, D., \& {Spitkovsky}, A. 2014, \apj, 783, 91,
  \dodoi{10.1088/0004-637X/783/2/91}

\bibitem[{{Chevalier}(1982{\natexlab{a}})}]{Chevalier82a}
{Chevalier}, R.~A. 1982{\natexlab{a}}, \apj, 258, 790, \dodoi{10.1086/160126}

\bibitem[{{Chevalier}(1982{\natexlab{b}})}]{Chevalier82b}
---. 1982{\natexlab{b}}, \apj, 259, 302

\bibitem[{{Chevalier}(1998)}]{Chevalier98}
---. 1998, \apj, 499, 810, \dodoi{10.1086/305676}

\bibitem[{{Chevalier} \& {Fransson}(1994)}]{ChevalierFransson94}
{Chevalier}, R.~A., \& {Fransson}, C. 1994, \apj, 420, 268,
  \dodoi{10.1086/173557}

\bibitem[{{Chevalier} \& {Fransson}(2006)}]{ChevalierFransson06}
---. 2006, \apj, 651, 381, \dodoi{10.1086/507606}

\bibitem[{{Chevalier} {et~al.}(2004){Chevalier}, {Li}, \&
  {Fransson}}]{Chevalier+04}
{Chevalier}, R.~A., {Li}, Z.-Y., \& {Fransson}, C. 2004, \apj, 606, 369,
  \dodoi{10.1086/382867}

\bibitem[{{Chomiuk} \& {Wilcots}(2009)}]{Chomiuk09}
{Chomiuk}, L., \& {Wilcots}, E.~M. 2009, \apj, 703, 370,
  \dodoi{10.1088/0004-637X/703/1/370}

\bibitem[{{Chomiuk} {et~al.}(2012){Chomiuk}, {Soderberg}, {Moe}, {Chevalier},
  {Rupen}, {Badenes}, {Margutti}, {Fransson}, {Fong}, \&
  {Dittmann}}]{Chomiuk2012}
{Chomiuk}, L., {Soderberg}, A.~M., {Moe}, M., {et~al.} 2012, \apj, 750, 164,
  \dodoi{10.1088/0004-637X/750/2/164}

\bibitem[{{Chomiuk} {et~al.}(2016){Chomiuk}, {Soderberg}, {Chevalier},
  {Bruzewski}, {Foley}, {Parrent}, {Strader}, {Badenes}, {Fransson}, {Kamble},
  {Margutti}, {Rupen}, \& {Simon}}]{Chomiuk+16}
{Chomiuk}, L., {Soderberg}, A.~M., {Chevalier}, R.~A., {et~al.} 2016, \apj,
  821, 119, \dodoi{10.3847/0004-637X/821/2/119}

\bibitem[{{Churazov} {et~al.}(2015){Churazov}, {Sunyaev}, {Isern}, {Bikmaev},
  {Bravo}, {Chugai}, {Grebenev}, {Jean}, {Kn{\"o}dlseder}, {Lebrun}, \&
  {Kuulkers}}]{Churazov+15}
{Churazov}, E., {Sunyaev}, R., {Isern}, J., {et~al.} 2015, \apj, 812, 62

\bibitem[{{Crumley} {et~al.}(2019){Crumley}, {Caprioli}, {Markoff}, \&
  {Spitkovsky}}]{Crumley+19}
{Crumley}, P., {Caprioli}, D., {Markoff}, S., \& {Spitkovsky}, A. 2019, \mnras,
  485, 5105, \dodoi{10.1093/mnras/stz232}

\bibitem[{{Deckers} {et~al.}(2022){Deckers}, {Maguire}, {Magee}, {Dimitriadis},
  {Smith}, {Sainz de Murieta}, {Miller}, {Goobar}, {Nordin}, {Rigault},
  {Bellm}, {Coughlin}, {Laher}, {Shupe}, {Graham}, {Kasliwal}, \&
  {Walters}}]{Deckers+22}
{Deckers}, M., {Maguire}, K., {Magee}, M.~R., {et~al.} 2022, arXiv e-prints,
  arXiv:2202.12914

\bibitem[{{Diesing} \& {Caprioli}(2021)}]{DiesingCaprioli21}
{Diesing}, R., \& {Caprioli}, D. 2021, \apj, 922, 1

\bibitem[{{Duffell} \& {Kasen}(2017)}]{DuffellKasen17}
{Duffell}, P.~C., \& {Kasen}, D. 2017, \apj, 842, 18

\bibitem[{{Freedman} {et~al.}(2019){Freedman}, {Madore}, {Hatt}, {Hoyt},
  {Jang}, {Beaton}, {Burns}, {Lee}, {Monson}, {Neeley}, {Phillips}, {Rich}, \&
  {Seibert}}]{Freedman+19}
{Freedman}, W.~L., {Madore}, B.~F., {Hatt}, D., {et~al.} 2019, \apj, 882, 34

\bibitem[{{Green}(2019)}]{Green+19}
{Green}, D.~A. 2019, Journal of Astrophysics and Astronomy, 40, 36,
  \dodoi{10.1007/s12036-019-9601-6}

\bibitem[{{Hachisu} {et~al.}(1996){Hachisu}, {Kato}, \& {Nomoto}}]{Hachisu+96}
{Hachisu}, I., {Kato}, M., \& {Nomoto}, K. 1996, \apjl, 470, L97

\bibitem[{{Hachisu} {et~al.}(1999){Hachisu}, {Kato}, {Nomoto}, \&
  {Umeda}}]{Hachisu+99}
{Hachisu}, I., {Kato}, M., {Nomoto}, K., \& {Umeda}, H. 1999, \apj, 519, 314

\bibitem[{{Hamuy} {et~al.}(2003){Hamuy}, {Phillips}, {Suntzeff}, {Maza},
  {Gonz{\'a}lez}, {Roth}, {Krisciunas}, {Morrell}, {Green}, {Persson}, \&
  {McCarthy}}]{Hamuy+03}
{Hamuy}, M., {Phillips}, M.~M., {Suntzeff}, N.~B., {et~al.} 2003, \nat, 424,
  651, \dodoi{10.1038/nature01854}

\bibitem[{{Harris} {et~al.}(2016){Harris}, {Nugent}, \& {Kasen}}]{HNK16}
{Harris}, C.~E., {Nugent}, P.~E., \& {Kasen}, D.~N. 2016, \apj, 823, 100,
  \dodoi{10.3847/0004-637X/823/2/100}

\bibitem[{{Horesh} {et~al.}(2013){Horesh}, {Stockdale}, {Fox}, {Frail},
  {Carpenter}, {Kulkarni}, {Ofek}, {Gal-Yam}, {Kasliwal}, {Arcavi}, {Quimby},
  {Cenko}, {Nugent}, {Bloom}, {Law}, {Poznanski}, {Gorbikov}, {Polishook},
  {Yaron}, {Ryder}, {Weiler}, {Bauer}, {Van Dyk}, {Immler}, {Panagia},
  {Pooley}, \& {Kassim}}]{Horesh+13}
{Horesh}, A., {Stockdale}, C., {Fox}, D.~B., {et~al.} 2013, \mnras, 436, 1258

\bibitem[{{Hosseinzadeh} {et~al.}(2017){Hosseinzadeh}, {Sand}, {Valenti},
  {Brown}, {Howell}, {McCully}, {Kasen}, {Arcavi}, {Bostroem}, {Tartaglia},
  {Hsiao}, {Davis}, {Shahbandeh}, \& {Stritzinger}}]{Hosseinzadeh+17}
{Hosseinzadeh}, G., {Sand}, D.~J., {Valenti}, S., {et~al.} 2017, \apjl, 845,
  L11, \dodoi{10.3847/2041-8213/aa8402}

\bibitem[{{Hosseinzadeh} {et~al.}(2022){Hosseinzadeh}, {Sand}, {Lundqvist},
  {Andrews}, {Bostroem}, {Dong}, {Janzen}, {Jencson}, {Lundquist}, {Meza},
  {Pearson}, {Valenti}, {Wyatt}, {Burke}, {Howell}, {McCully}, {Newsome},
  {Padilla Gonzalez}, {Pellegrino}, {Terreran}, {Kwok}, {Jha}, {Strader},
  {Kundu}, {Ryder}, {Haislip}, {Kouprianov}, \& {Reichart}}]{Hosseinzadeh+22}
{Hosseinzadeh}, G., {Sand}, D.~J., {Lundqvist}, P., {et~al.} 2022, arXiv
  e-prints, arXiv:2205.02236

\bibitem[{{Huang} \& {Yu}(1996)}]{HuangYu96}
{Huang}, R.-q., \& {Yu}, K.~N. 1996, \caa, 20, 175

\bibitem[{{Hunter}(2007)}]{Matplotlib}
{Hunter}, J.~D. 2007, Computing in Science \& Engineering, 9, 90

\bibitem[{{Iben} \& {Tutukov}(1984)}]{IbenTutukov84}
{Iben}, I., J., \& {Tutukov}, A.~V. 1984, \apjs, 54, 335

\bibitem[{{Jacobson-Gal{\'a}n} {et~al.}(2020){Jacobson-Gal{\'a}n}, {Margutti},
  {Kilpatrick}, {Hiramatsu}, {Perets}, {Khatami}, {Foley}, {Raymond}, {Yoon},
  {Bobrick}, {Zenati}, {Galbany}, {Andrews}, {Brown}, {Cartier}, {Coppejans},
  {Dimitriadis}, {Dobson}, {Hajela}, {Howell}, {Kuncarayakti}, {Milisavljevic},
  {Rahman}, {Rojas-Bravo}, {Sand}, {Shepherd}, {Smartt}, {Stacey}, {Stroh},
  {Swift}, {Terreran}, {Vinko}, {Wang}, {Anderson}, {Baron}, {Berger},
  {Blanchard}, {Burke}, {Coulter}, {DeMarchi}, {DerKacy}, {Fremling}, {Gomez},
  {Gromadzki}, {Hosseinzadeh}, {Kasen}, {Kriskovics}, {McCully},
  {M{\"u}ller-Bravo}, {Nicholl}, {Ordasi}, {Pellegrino}, {Piro}, {P{\'a}l},
  {Ren}, {Rest}, {Rich}, {Sai}, {S{\'a}rneczky}, {Shen}, {Short}, {Siebert},
  {Stauffer}, {Szak{\'a}ts}, {Zhang}, {Zhang}, \& {Zhang}}]{Jacobson-Galan+20}
{Jacobson-Gal{\'a}n}, W.~V., {Margutti}, R., {Kilpatrick}, C.~D., {et~al.}
  2020, \apj, 898, 166

\bibitem[{{Jiang} {et~al.}(2021){Jiang}, {Maeda}, {Kawabata}, {Doi},
  {Shigeyama}, {Tanaka}, {Tominaga}, {Nomoto}, {Niino}, {Sako}, {Ohsawa},
  {Schramm}, {Yamanaka}, {Kobayashi}, {Takahashi}, {Nakaoka}, {Kawabata},
  {Isogai}, {Aoki}, {Kondo}, {Mori}, {Arimatsu}, {Kasuga}, {Okumura},
  {Urakawa}, {Reichart}, {Taguchi}, {Arima}, {Beniyama}, {Uno}, \&
  {Hamada}}]{Jiang+21}
{Jiang}, J.-a., {Maeda}, K., {Kawabata}, M., {et~al.} 2021, \apjl, 923, L8

\bibitem[{{Jones} {et~al.}(2001){Jones}, {Oliphant}, {Peterson},
  {et~al.}}]{SciPy}
{Jones}, E., {Oliphant}, T., {Peterson}, P., {et~al.} 2001, {SciPy}: Open
  source scientific tools for {Python}

\bibitem[{{Jord{\'a}n} {et~al.}(2005){Jord{\'a}n}, {C{\^o}t{\'e}}, {Blakeslee},
  {Ferrarese}, {McLaughlin}, {Mei}, {Peng}, {Tonry}, {Merritt},
  {Milosavljevi{\'c}}, {Sarazin}, {Sivakoff}, \& {West}}]{Jordan+05}
{Jord{\'a}n}, A., {C{\^o}t{\'e}}, P., {Blakeslee}, J.~P., {et~al.} 2005, \apj,
  634, 1002, \dodoi{10.1086/497092}

\bibitem[{{Jord{\'a}n} {et~al.}(2007){Jord{\'a}n}, {McLaughlin},
  {C{\^o}t{\'e}}, {Ferrarese}, {Peng}, {Mei}, {Villegas}, {Merritt}, {Tonry},
  \& {West}}]{Jordan+07}
{Jord{\'a}n}, A., {McLaughlin}, D.~E., {C{\^o}t{\'e}}, P., {et~al.} 2007,
  \apjs, 171, 101, \dodoi{10.1086/516840}

\bibitem[{{Kasen}(2010)}]{Kasen10}
{Kasen}, D. 2010, \apj, 708, 1025, \dodoi{10.1088/0004-637X/708/2/1025}

\bibitem[{{Li} {et~al.}(2011){Li}, {Bloom}, {Podsiadlowski}, {Miller}, {Cenko},
  {Jha}, {Sullivan}, {Howell}, {Nugent}, {Butler}, {Ofek}, {Kasliwal},
  {Richards}, {Stockton}, {Shih}, {Bildsten}, {Shara}, {Bibby}, {Filippenko},
  {Ganeshalingam}, {Silverman}, {Kulkarni}, {Law}, {Poznanski}, {Quimby},
  {McCully}, {Patel}, {Maguire}, \& {Shen}}]{Li+11c}
{Li}, W., {Bloom}, J.~S., {Podsiadlowski}, P., {et~al.} 2011, \nat, 480, 348

\bibitem[{{Livio} \& {Mazzali}(2018)}]{LivioMazzali18}
{Livio}, M., \& {Mazzali}, P. 2018, \physrep, 736, 1

\bibitem[{{Lundqvist} {et~al.}(2020){Lundqvist}, {Kundu}, {P{\'e}rez-Torres},
  {Ryder}, {Bj{\"o}rnsson}, {Moldon}, {Argo}, {Beswick}, {Alberdi}, \&
  {Kool}}]{Lundqvist+20}
{Lundqvist}, P., {Kundu}, E., {P{\'e}rez-Torres}, M.~A., {et~al.} 2020, \apj,
  890, 159

\bibitem[{{Macaulay} {et~al.}(2019){Macaulay}, {Nichol}, {Bacon}, {Brout},
  {Davis}, {Zhang}, {Bassett}, {Scolnic}, {M{\"o}ller}, {D'Andrea}, {Hinton},
  {Kessler}, {Kim}, {Lasker}, {Lidman}, {Sako}, {Smith}, {Sullivan}, {Abbott},
  {Allam}, {Annis}, {Asorey}, {Avila}, {Bechtol}, {Brooks}, {Brown}, {Burke},
  {Calcino}, {Carnero Rosell}, {Carollo}, {Carrasco Kind}, {Carretero},
  {Castander}, {Collett}, {Crocce}, {Cunha}, {da Costa}, {Davis}, {De Vicente},
  {Diehl}, {Doel}, {Drlica-Wagner}, {Eifler}, {Estrada}, {Evrard},
  {Filippenko}, {Finley}, {Flaugher}, {Foley}, {Fosalba}, {Frieman}, {Galbany},
  {Garc{\'\i}a-Bellido}, {Gaztanaga}, {Glazebrook}, {Gonz{\'a}lez-Gait{\'a}n},
  {Gruen}, {Gruendl}, {Gschwend}, {Gutierrez}, {Hartley}, {Hollowood},
  {Honscheid}, {Hoormann}, {Hoyle}, {Huterer}, {Jain}, {James}, {Jeltema},
  {Kasai}, {Krause}, {Kuehn}, {Kuropatkin}, {Lahav}, {Lewis}, {Li}, {Lima},
  {Lin}, {Maia}, {Marshall}, {Martini}, {Miquel}, {Nugent}, {Palmese}, {Pan},
  {Plazas}, {Romer}, {Roodman}, {Sanchez}, {Scarpine}, {Schindler},
  {Schubnell}, {Serrano}, {Sevilla-Noarbe}, {Sharp}, {Soares-Santos},
  {Sobreira}, {Sommer}, {Suchyta}, {Swann}, {Swanson}, {Tarle}, {Thomas},
  {Thomas}, {Tucker}, {Uddin}, {Vikram}, {Walker}, {Wiseman}, \& {DES
  Collaboration}}]{DES19b}
{Macaulay}, E., {Nichol}, R.~C., {Bacon}, D., {et~al.} 2019, \mnras, 486, 2184

\bibitem[{{Magee} \& {Maguire}(2020)}]{Magee2020}
{Magee}, M.~R., \& {Maguire}, K. 2020, \aap, 642, A189,
  \dodoi{10.1051/0004-6361/202037870}

\bibitem[{{Maoz} {et~al.}(2014){Maoz}, {Mannucci}, \& {Nelemans}}]{Maoz+14}
{Maoz}, D., {Mannucci}, F., \& {Nelemans}, G. 2014, \araa, 52, 107

\bibitem[{{Marcowith} {et~al.}(2020){Marcowith}, {Ferrand}, {Grech}, {Meliani},
  {Plotnikov}, \& {Walder}}]{Marcowith+20}
{Marcowith}, A., {Ferrand}, G., {Grech}, M., {et~al.} 2020, Living Reviews in
  Computational Astrophysics, 6, 1, \dodoi{10.1007/s41115-020-0007-6}

\bibitem[{{Margutti} {et~al.}(2014){Margutti}, {Parrent}, {Kamble},
  {Soderberg}, {Foley}, {Milisavljevic}, {Drout}, \& {Kirshner}}]{Margutti+14}
{Margutti}, R., {Parrent}, J., {Kamble}, A., {et~al.} 2014, \apj, 790, 52

\bibitem[{{Margutti} {et~al.}(2012){Margutti}, {Soderberg}, {Chomiuk},
  {Chevalier}, {Hurley}, {Milisavljevic}, {Foley}, {Hughes}, {Slane},
  {Fransson}, {Moe}, {Barthelmy}, {Boynton}, {Briggs}, {Connaughton}, {Costa},
  {Cummings}, {Del Monte}, {Enos}, {Fellows}, {Feroci}, {Fukazawa}, {Gehrels},
  {Goldsten}, {Golovin}, {Hanabata}, {Harshman}, {Krimm}, {Litvak},
  {Makishima}, {Marisaldi}, {Mitrofanov}, {Murakami}, {Ohno}, {Palmer},
  {Sanin}, {Starr}, {Svinkin}, {Takahashi}, {Tashiro}, {Terada}, \&
  {Yamaoka}}]{Margutti+12}
{Margutti}, R., {Soderberg}, A.~M., {Chomiuk}, L., {et~al.} 2012, \apj, 751,
  134

\bibitem[{{Marion} {et~al.}(2016){Marion}, {Brown}, {Vink{\'o}}, {Silverman},
  {Sand}, {Challis}, {Kirshner}, {Wheeler}, {Berlind}, {Brown}, {Calkins},
  {Camacho}, {Dhungana}, {Foley}, {Friedman}, {Graham}, {Howell}, {Hsiao},
  {Irwin}, {Jha}, {Kehoe}, {Macri}, {Maeda}, {Mandel}, {McCully}, {Pandya},
  {Rines}, {Wilhelmy}, \& {Zheng}}]{Marion+16}
{Marion}, G.~H., {Brown}, P.~J., {Vink{\'o}}, J., {et~al.} 2016, \apj, 820, 92,
  \dodoi{10.3847/0004-637X/820/2/92}

\bibitem[{{Marongiu} {et~al.}(2022){Marongiu}, {Guidorzi}, {Stratta}, {Gomboc},
  {Jordana-Mitjans}, {Dichiara}, {Kobayashi}, {Kopa{\v{c}}}, \&
  {Mundell}}]{Marongiu+22}
{Marongiu}, M., {Guidorzi}, C., {Stratta}, G., {et~al.} 2022, \aap, 658, A11

\bibitem[{{McCully} {et~al.}(2014){McCully}, {Jha}, {Foley}, {Bildsten},
  {Fong}, {Kirshner}, {Marion}, {Riess}, \& {Stritzinger}}]{McCully+14}
{McCully}, C., {Jha}, S.~W., {Foley}, R.~J., {et~al.} 2014, \nat, 512, 54

\bibitem[{{Miller} {et~al.}(2018){Miller}, {Cao}, {Piro}, {Blagorodnova},
  {Bue}, {Cenko}, {Dhawan}, {Ferretti}, {Fox}, {Fremling}, {Goobar}, {Howell},
  {Hosseinzadeh}, {Kasliwal}, {Laher}, {Lunnan}, {Masci}, {McCully}, {Nugent},
  {Sollerman}, {Taddia}, \& {Kulkarni}}]{Miller+18}
{Miller}, A.~A., {Cao}, Y., {Piro}, A.~L., {et~al.} 2018, \apj, 852, 100,
  \dodoi{10.3847/1538-4357/aaa01f}

\bibitem[{{Moore} \& {Bildsten}(2012)}]{MooreBildsten12}
{Moore}, K., \& {Bildsten}, L. 2012, \apj, 761, 182,
  \dodoi{10.1088/0004-637X/761/2/182}

\bibitem[{{Morozova} {et~al.}(2015){Morozova}, {Piro}, {Renzo}, {Ott},
  {Clausen}, {Couch}, {Ellis}, \& {Roberts}}]{Morozova+15}
{Morozova}, V., {Piro}, A.~L., {Renzo}, M., {et~al.} 2015, \apj, 814, 63

\bibitem[{{Nakar} \& {Sari}(2010)}]{NakarSari10}
{Nakar}, E., \& {Sari}, R. 2010, \apj, 725, 904

\bibitem[{{Ofek} {et~al.}(2013){Ofek}, {Lin}, {Kouveliotou}, {Younes},
  {G{\"o}{\v g}{\"u}{\c s}}, {Kasliwal}, \& {Cao}}]{Ofek+13}
{Ofek}, E.~O., {Lin}, L., {Kouveliotou}, C., {et~al.} 2013, \apj, 768, 47

\bibitem[{{Oliphant}(2006)}]{NumPy}
{Oliphant}, T. 2006, {A guide to NumPy} (USA: Trelgol Publishing)

\bibitem[{{Pakmor} {et~al.}(2012){Pakmor}, {Kromer}, {Taubenberger}, {Sim},
  {R{\"o}pke}, \& {Hillebrandt}}]{Pakmor+12}
{Pakmor}, R., {Kromer}, M., {Taubenberger}, S., {et~al.} 2012, \apjl, 747, L10,
  \dodoi{10.1088/2041-8205/747/1/L10}

\bibitem[{{Panaitescu} \& {Kumar}(2002)}]{PanaitescuKumar02}
{Panaitescu}, A., \& {Kumar}, P. 2002, \apj, 571, 779

\bibitem[{{Paxton} {et~al.}(2011){Paxton}, {Bildsten}, {Dotter}, {Herwig},
  {Lesaffre}, \& {Timmes}}]{Paxton+11}
{Paxton}, B., {Bildsten}, L., {Dotter}, A., {et~al.} 2011, \apjs, 192, 3

\bibitem[{{Pellegrino} {et~al.}(2020){Pellegrino}, {Howell}, {Sarbadhicary},
  {Burke}, {Hiramatsu}, {McCully}, {Milne}, {Andrews}, {Brown}, {Chomiuk},
  {Hsiao}, {Sand}, {Shahbandeh}, {Smith}, {Valenti}, {Vink{\'o}}, {Wheeler},
  {Wyatt}, \& {Yang}}]{Pellegrino2020}
{Pellegrino}, C., {Howell}, D.~A., {Sarbadhicary}, S.~K., {et~al.} 2020, \apj,
  897, 159, \dodoi{10.3847/1538-4357/ab8e3f}

\bibitem[{{P{\'e}rez-Torres} {et~al.}(2014){P{\'e}rez-Torres}, {Lundqvist},
  {Beswick}, {Bj{\"o}rnsson}, {Muxlow}, {Paragi}, {Ryder}, {Alberdi},
  {Fransson}, {Marcaide}, {Mart{\'{\i}}-Vidal}, {Ros}, {Argo}, \&
  {Guirado}}]{PerezTorres+14}
{P{\'e}rez-Torres}, M.~A., {Lundqvist}, P., {Beswick}, R.~J., {et~al.} 2014,
  \apj, 792, 38

\bibitem[{{Piro} \& {Morozova}(2016)}]{PiroMorozova16}
{Piro}, A.~L., \& {Morozova}, V.~S. 2016, \apj, 826, 96

\bibitem[{{Polin} {et~al.}(2019){Polin}, {Nugent}, \& {Kasen}}]{Polin2019}
{Polin}, A., {Nugent}, P., \& {Kasen}, D. 2019, \apj, 873, 84,
  \dodoi{10.3847/1538-4357/aafb6a}

\bibitem[{{Raskin} \& {Kasen}(2013)}]{RaskinKasen13}
{Raskin}, C., \& {Kasen}, D. 2013, \apj, 772, 1,
  \dodoi{10.1088/0004-637X/772/1/1}

\bibitem[{{Reynolds} {et~al.}(2021){Reynolds}, {Williams}, {Borkowski}, \&
  {Long}}]{Reynolds+21}
{Reynolds}, S.~P., {Williams}, B.~J., {Borkowski}, K.~J., \& {Long}, K.~S.
  2021, \apj, 917, 55, \dodoi{10.3847/1538-4357/ac0ced}

\bibitem[{{Riess} {et~al.}(2016){Riess}, {Macri}, {Hoffmann}, {Scolnic},
  {Casertano}, {Filippenko}, {Tucker}, {Reid}, {Jones}, {Silverman},
  {Chornock}, {Challis}, {Yuan}, {Brown}, \& {Foley}}]{Riess+16}
{Riess}, A.~G., {Macri}, L.~M., {Hoffmann}, S.~L., {et~al.} 2016, \apj, 826, 56

\bibitem[{{Sai} {et~al.}(2022){Sai}, {Wang}, {Elias-Rosa}, {Yang}, {Zhang},
  {Fiore}, {Hsiao}, {Isern}, {Itagaki}, {Li}, {Li}, {Pessi}, {Phillips},
  {Schuldt}, {Shahbandeh}, {Stritzinger}, {Tomasella}, {Vogl}, {Wang}, {Wang},
  {Wu}, {Yang}, {Zhang}, {Zhang}, \& {Zhang}}]{Sai+22}
{Sai}, H., {Wang}, X., {Elias-Rosa}, N., {et~al.} 2022, arXiv e-prints,
  arXiv:2205.15596

\bibitem[{{Schwab} {et~al.}(2012){Schwab}, {Shen}, {Quataert}, {Dan}, \&
  {Rosswog}}]{Schwab+12}
{Schwab}, J., {Shen}, K.~J., {Quataert}, E., {Dan}, M., \& {Rosswog}, S. 2012,
  \mnras, 427, 190

\bibitem[{{Seaquist} \& {Taylor}(1990)}]{SeaquistTaylor90}
{Seaquist}, E.~R., \& {Taylor}, A.~R. 1990, \apj, 349, 313,
  \dodoi{10.1086/168315}

\bibitem[{{Shen} \& {Bildsten}(2007)}]{ShenBildsten07}
{Shen}, K.~J., \& {Bildsten}, L. 2007, \apj, 660, 1444

\bibitem[{{Shen} {et~al.}(2012){Shen}, {Bildsten}, {Kasen}, \&
  {Quataert}}]{Shen+12}
{Shen}, K.~J., {Bildsten}, L., {Kasen}, D., \& {Quataert}, E. 2012, \apj, 748,
  35

\bibitem[{{Silverman} {et~al.}(2013){Silverman}, {Nugent}, {Gal-Yam},
  {Sullivan}, {Howell}, {Filippenko}, {Arcavi}, {Ben-Ami}, {Bloom}, {Cenko},
  {Cao}, {Chornock}, {Clubb}, {Coil}, {Foley}, {Graham}, {Griffith}, {Horesh},
  {Kasliwal}, {Kulkarni}, {Leonard}, {Li}, {Matheson}, {Miller}, {Modjaz},
  {Ofek}, {Pan}, {Perley}, {Poznanski}, {Quimby}, {Steele}, {Sternberg}, {Xu},
  \& {Yaron}}]{Silverman+13}
{Silverman}, J.~M., {Nugent}, P.~E., {Gal-Yam}, A., {et~al.} 2013, \apjs, 207,
  3

\bibitem[{{Soderberg} {et~al.}(2005){Soderberg}, {Kulkarni}, {Berger},
  {Chevalier}, {Frail}, {Fox}, \& {Walker}}]{Soderberg+05}
{Soderberg}, A.~M., {Kulkarni}, S.~R., {Berger}, E., {et~al.} 2005, \apj, 621,
  908

\bibitem[{{Soker}(2013)}]{Soker13}
{Soker}, N. 2013, in IAU Symposium, Vol. 281, Binary Paths to Type Ia
  Supernovae Explosions, ed. R.~{Di Stefano}, M.~{Orio}, \& M.~{Moe}, 72--75,
  \dodoi{10.1017/S174392131201472X}

\bibitem[{{Sorce} {et~al.}(2014){Sorce}, {Tully}, {Courtois}, {Jarrett},
  {Neill}, \& {Shaya}}]{Sorce+14}
{Sorce}, J.~G., {Tully}, R.~B., {Courtois}, H.~M., {et~al.} 2014, \mnras, 444,
  527, \dodoi{10.1093/mnras/stu1450}

\bibitem[{{Springob} {et~al.}(2009){Springob}, {Masters}, {Haynes},
  {Giovanelli}, \& {Marinoni}}]{Springob+09}
{Springob}, C.~M., {Masters}, K.~L., {Haynes}, M.~P., {Giovanelli}, R., \&
  {Marinoni}, C. 2009, \apjs, 182, 474, \dodoi{10.1088/0067-0049/182/1/474}

\bibitem[{{Theureau} {et~al.}(2007){Theureau}, {Hanski}, {Coudreau}, {Hallet},
  \& {Martin}}]{Theureau+07}
{Theureau}, G., {Hanski}, M.~O., {Coudreau}, N., {Hallet}, N., \& {Martin},
  J.~M. 2007, \aap, 465, 71, \dodoi{10.1051/0004-6361:20066187}

\bibitem[{{Tully} {et~al.}(2016){Tully}, {Courtois}, \& {Sorce}}]{Tully+16}
{Tully}, R.~B., {Courtois}, H.~M., \& {Sorce}, J.~G. 2016, \aj, 152, 50,
  \dodoi{10.3847/0004-6256/152/2/50}

\bibitem[{{Tully} {et~al.}(2009){Tully}, {Rizzi}, {Shaya}, {Courtois},
  {Makarov}, \& {Jacobs}}]{Tully+09}
{Tully}, R.~B., {Rizzi}, L., {Shaya}, E.~J., {et~al.} 2009, \aj, 138, 323,
  \dodoi{10.1088/0004-6256/138/2/323}

\bibitem[{{Tully} {et~al.}(2013){Tully}, {Courtois}, {Dolphin}, {Fisher},
  {H{\'e}raudeau}, {Jacobs}, {Karachentsev}, {Makarov}, {Makarova},
  {Mitronova}, {Rizzi}, {Shaya}, {Sorce}, \& {Wu}}]{Tully+13}
{Tully}, R.~B., {Courtois}, H.~M., {Dolphin}, A.~E., {et~al.} 2013, \aj, 146,
  86, \dodoi{10.1088/0004-6256/146/4/86}

\bibitem[{{Villegas} {et~al.}(2010){Villegas}, {Jord{\'a}n}, {Peng},
  {Blakeslee}, {C{\^o}t{\'e}}, {Ferrarese}, {Kissler-Patig}, {Mei}, {Infante},
  {Tonry}, \& {West}}]{Villegas+10}
{Villegas}, D., {Jord{\'a}n}, A., {Peng}, E.~W., {et~al.} 2010, \apj, 717, 603,
  \dodoi{10.1088/0004-637X/717/2/603}

\bibitem[{{Weiler} {et~al.}(1986){Weiler}, {Sramek}, {Panagia}, {van der
  Hulst}, \& {Salvati}}]{Weiler+86}
{Weiler}, K.~W., {Sramek}, R.~A., {Panagia}, N., {van der Hulst}, J.~M., \&
  {Salvati}, M. 1986, \apj, 301, 790

\bibitem[{{Whelan} \& {Iben}(1973)}]{WhelanIben73}
{Whelan}, J., \& {Iben}, Icko, J. 1973, \apj, 186, 1007

\bibitem[{{Willick} {et~al.}(1997){Willick}, {Courteau}, {Faber}, {Burstein},
  {Dekel}, \& {Strauss}}]{Willick+97}
{Willick}, J.~A., {Courteau}, S., {Faber}, S.~M., {et~al.} 1997, \apjs, 109,
  333, \dodoi{10.1086/312983}

\bibitem[{{Wright}(2006)}]{cosmologycalculator}
{Wright}, E.~L. 2006, \pasp, 118, 1711, \dodoi{10.1086/510102}

\bibitem[{{Yost} {et~al.}(2003){Yost}, {Harrison}, {Sari}, \&
  {Frail}}]{Yost+03}
{Yost}, S.~A., {Harrison}, F.~A., {Sari}, R., \& {Frail}, D.~A. 2003, \apj,
  597, 459, \dodoi{10.1086/378288}

\end{thebibliography}
\bibliographystyle{aasjournal}

\end{document}